\documentstyle[galley,epsfig]{mn2e}
\title{Chemical  analysis of  CH stars - II: atmospheric parameters and elemental abundances}
\title[Chemical  analysis of  CH stars - II: atmospheric parameters and elemental abundances]
{Chemical analysis of  CH stars - II: atmospheric parameters and elemental abundances}

\author[Drisya Karinkuzhi et al.]{Drisya Karinkuzhi$^{1,2}$, Aruna Goswami$^{1}$  \\
    $^{1}$Indian Institute of Astrophysics, Koramangala, Bangalore 560034,
    India; drisya@iiap.res.in, aruna@iiap.res.in\\ $^{2}$ Department of physics, Bangalore university, Jnana Bharathi Campus, Karntaka 560056, India\\
}    
\begin{document}

\date{ Accepted ---;  Received ---;  in original 
form --- \large \bf }

\pagerange{\pageref{firstpage}--\pageref{lastpage}} \pubyear{2014}

\maketitle

\label{firstpage}

\begin{abstract}
We present detailed chemical analyses for a sample of twelve  stars 
selected from the CH star catalogue of Bartkevicius (1996). The sample 
includes two confirmed binaries, four objects that are known to show 
radial velocity variations and the rest with no information on the
binary status. A primary objective is to examine if all these objects
exhibit  chemical abundances  characteristics of CH stars, based on 
detailed  chemical composition study  using  high  resolution spectra. 
  We have used high resolution (R ${\sim}$ 42000) spectra  from
the ELODIE archive. These spectra cover 3900 ${\rm \AA}$ to 6800 ${\rm \AA}$ 
in the wavelength range. We have estimated the stellar atmospheric parameters, 
the effective temperature $T_{eff}$, the surface gravity log g, and 
metallicity [Fe/H] from LTE analysis using model atmospheres.
Estimated temperatures of these objects cover a wide range from 4200 K to 
6640 K, the surface gravity from 0.6 to 4.3 and metallicity from $-$0.13 
to $-$1.5. We report updates on elemental abundances for several heavy 
elements, Sr, Y, Zr, Ba, La, Ce, Pr, Nd, Sm, Eu and Dy. 
 For the object HD~89668 we present the first abundance analyses results. 
Enhancement of heavy elements relative to Fe, a  characteristic property of 
CH stars is  evident from our analyses in case of  four objects, HD~92545, 
HD~104979, HD~107574 and HD~204613. A parametric 
model based study is performed to understand the relative contributions 
from the s- and r-process to the abundances of the heavy elements.

\end{abstract}


\section{Introduction}
CH stars are characterized by iron deficiency and enhancement of Carbon 
and s-process elements. Majority of the CH stars are known as binaries 
with  white dwarf companions that are presently not visible 
(McClure 1983, 1984, McClure \& Woodsworth 1990). The companion white dwarfs 
produced heavy elements while passing through the AGB stage of evolution; 
these material are  received by the CH stars through mass transfer
enriching their surface chemical composition.
CH stars thus provide an important means to study  the production and 
distribution of heavy elements arising from AGB nucleosynthesis.

Inspite of their usefulness, literature survey reveals that detailed
chemical composition studies are not available for many CH stars.
The CH star catalogue of Bartkevicius (1996) lists about 261 objects, 
 17  of which  belong to ${\omega}$ Cen globular cluster. 
Many of the objects  listed in this  catalogue   have no information 
on binary status.
 
It would be interesting to compare and examine the abundance patterns 
of elements observed  in the confirmed binaries  with their counterparts 
in  objects that  have no  information on binary status. While long-term 
radial velocity monitoring are expected to throw light on the binary status, 
detailed chemical composition studies could also reflect on the 
binary origin.  

Previous studies along this line include a detailed chemical composition 
study of ten objects from the Bartkevicius catalogue by Karinkuzhi \& Goswami (2014) 
(paper 1). This study revealed that  only five objects out of ten, exhibit 
 abundances of heavy elements with [Ba/Fe] ${\textgreater}$ 1, a
characteristic of CH stars.  Four objects show either near-solar values 
or [Ba/Fe] ${\textless}$ 0. The remaining one object, HD~4395 gave 
[Ba/Fe] ${\sim}$0.79. Based on their analyses the authors concluded that   
out of ten, only five  objects are bonafide  CH stars.   

As far as the chemical composition is concerned, CH stars 
(with $-$0.2 $<$[Fe/H] $<$ $-$2) and the class of carbon-enhanced
metal-poor (CEMP)-s   ([C/Fe] ${\textgreater}$ 1, [Fe/H] ${\textless}$ $-$2; 
Beers and Christlieb 2005) stars are believed 
to have  a similar origin. Medium resolution spectral analyses of about
300 faint high latitude carbon stars of Hamburg/ESO survey 
(Christlieb et al. 2001) have shown that  about 33 per cent of the objects are
 potential CH star candidates (Goswami 2005, Goswami et al. 2007, 2010a). 
Analyses of high resolution Subaru spectra for a sample  of them,
have shown the object  HE~1152$-$0355 to be a CH star, and  HE~1305+0007, 
a CEMP-r/s star (Goswami et al. 2006).
A large fraction of CEMP-s and CEMP-r/s stars show radial velocity 
variations, based on which
these stars are suggested to be all binaries (Lucatello et al. 2005),
and that the CEMP-s stars  are  the  more metal-poor 
counterparts of  CH stars.

Although high resolution spectroscopic analyses of CEMP stars have shown 
that a variety of production mechanisms are needed to explain the 
observed range of elemental abundance patterns in them, it is widely 
accepted that  the binary 
scenario of CH star formation  is the most likely formation mechanism 
also for CEMP-s  stars (Barbuy et al. 2005; Norris et al. 1997a, b, 2002;
Aoki et al.  2001, 2002a,b, 2007; Lucatello et al. 2005; Goswami 
et al. 2006, Goswami \& Aoki 2010b).

In this work we have considered another twelve objects from the catalogue of
  Bartkevicius (1996) for a detailed chemical composition study. 
Detailed high resolution spectroscopic analyses for this sample of  
objects are  either  not available in the literature or  limited  by 
resolution or wavelength range. 
Polarimetric studies of carbon stars by Goswami \& Karinkuzhi (2013)
include six objects from this sample.  Among these three objects
show percentage V-band polarization at a level ${\sim}$ 0.2 per cent  
(HD~55496 (p$_{v}$ per cent ${\sim}$ 0.18), HD~111721 (p$_{v}$ per cent ${\sim}$ 0.22),
and HD~164922 (p$_{v}$ per cent ${\sim}$ 0.28)) indicating presence of circumstellar
dust distribution in non-spherically symmetric envelopes. The other three 
objects, HD~92545, HD~107574 and HD~126681,  show V-band
percentage polarization at a level ${\textless}$ 0.1 per cent.  

The sample  of programme stars includes two confirmed binaries, HD~122202 and 
HD~204613.
Four objects in this sample,  HD~55496, HD~92545, HD~104979 and HD~107574 
are known to show radial velocity variability,  and for the rest, none of 
these two information is available. In the following text, for convenience,
we will refer the objects that are confirmed binaries as group one objects,
those with limited radial velocity information as group 2 objects and the
objects for which none of these  information are available as group three
objects.   One of our primary  objectives is to 
estimate the abundances of heavy elements  and critically  examine the  
abundance patterns and abundance ratios to check if they exhibit characteristic 
abundance patterns of CH stars.

Among CEMP stars the group of CEMP-r/s stars show enhancement of both
r- and s-process elements ( 0 $<$[Ba/Eu] $<$ 0.5 (Beers \& Christlieb 2005)).
None of our  objects in the  sample are found to  show [Ba/Eu] ratios 
in this range. Four objects show characteristic heavy element abundance 
patterns  of CH stars. Based on our analyses, the others certainly do not 
belong to this class of objects.

Source of the high resolution spectra is described in section 2. 
Estimates  of  radial velocities are presented in section 3. Temperature 
estimates from photometry are discussed in section 4. Estimation of 
stellar atmospheric parameters are presented  in section 5. Results 
of abundance analysis are discussed in section 6. In section 7 we 
present  brief discussions on each individual star.  Estimated stellar 
masses are discussed in section 8. A discussion on 
the parametric model based analysis is presented in section 9. 
Conclusions are drawn in section 10.

 \section {Spectra of the programme stars}
Low-resolution spectra of these objects obtained from 2m Himalayan Chandra 
Telescope at the Indian Astronomical Observatory (IAO), Hanle using HFOSC  
clearly show strong features due to carbon. HFOSC is an optical imager 
cum spectrograph for conducting low- and medium-resolution grism 
spectroscopy (http://www.iiap.res.in/iao/iao.html).
 High resolution spectra necessary for abundance analyses of the programme stars
 are  taken from the ELODIE archive (Moultaka et al. (2004)). 
This archive contains a large collection of high-resolution spectra 
acquired with the 1.93 m telescope at the Observatoire de  Haute Provence 
(OHP) using the ELODIE spectrograph (Baranne et al. 1996).  An online 
reduction software program TACOS  automatically performs optimal extraction 
and wavelength calibration of data.  The spectra consist of  67 orders 
with near constant inter order spacing. The  resolution of the spectra is
 ${\sim}$ 42000 and cover the wavelength range  
 3900 {\rm \AA} to 6800 {\rm \AA}. 
A few  sample spectra are shown in Figures 1 and 2.
The basic data  for  the programme stars obtained from the SIMBAD database 
are listed in Table 1. 
\begin{figure}
\centering
\includegraphics[height=8cm,width=8cm]{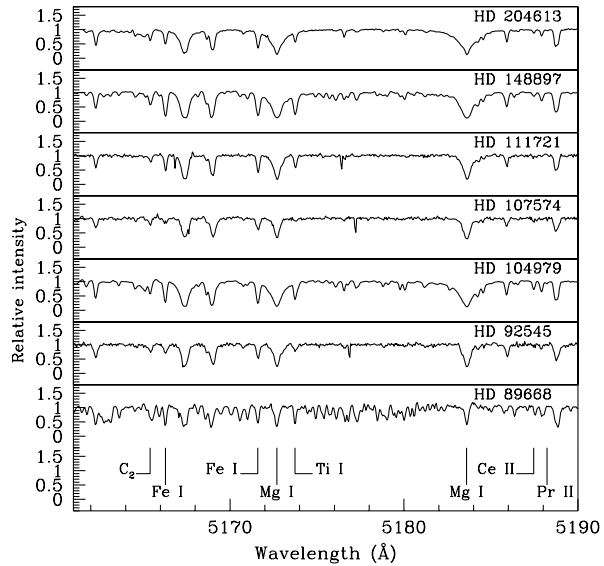}
\caption{Sample spectra of a few programme stars in the wavelength region 
5160 to 5190  {\rm \AA},}
\end{figure}
%\label (fig 1)

\begin{figure}
\centering
\includegraphics[height=8cm,width=8cm]{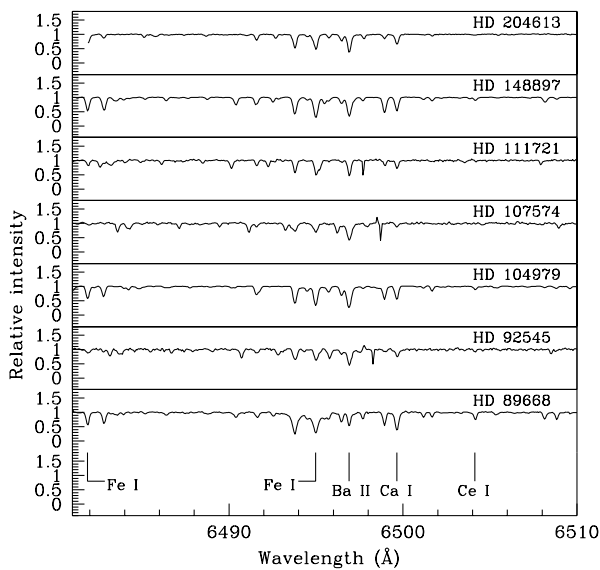}
\caption{ Spectra showing the wavelength region 6480 to 6510 
{\rm \AA}, for the same  stars as  in Figure 1.}
\end{figure}
\begin{table*}
{Table 1: Basic data for the programme stars}\\
\begin{tabular}{cccccccc}
\hline
Star Name.   & RA(2000) & DEC(2000)& B& V&J&H&K \\

  HD~55496    &	07 12 11.37& $-$22 59 00.61&	 9.30 &	 8.40 & 6.590&	 6.043	& 5.931\\
  HD~89668    &	10 20 43.40& $-$01 28 11.38&	10.50 &	 9.41 & 7.443&	 6.908	& 6.760\\
  HD~92545    &	10 40 57.70 &$-$12 11 44.23&	 9.07 &	 8.56 & 7.548&	 7.347	& 7.282\\
  HD~104979   &	12 05 12.54& +08 43 58.74&	 5.10 &	 4.13 & 2.459&	 1.987	& 1.869\\
  HD~107574   &	12 21 51.86& $-$18 24 00.15&	 8.99 &  8.54 &	 7.660&	 7.460	 &7.415\\
  HD~111721   &	12 51 25.19 &$-$13 29 28.17&	 8.78 &	 7.97 & 6.347&	 5.898	& 5.786\\
  HD~122202   &	14 00 18.96& +04 51 25.06&  	 9.85 &	 9.36 & 8.506&	 8.358	& 8.252\\
  HD~126681   &	14 27 24.91& $-$18 24 40.43&	 9.93 &	 9.32 & 8.044&	 7.709	& 7.631\\
  HD~148897   &	16 30 33.54& +20 28 45.07&	 6.50 &	 5.25 & 2.950&	 2.248&	 1.966\\
  HD~164922   &	18 02 30.86 &+26 18 46.80&	 7.79 &	 6.99 & 5.553&	 5.203	& 5.113\\
  HD~167768   &	18 16 53.10& $-$03 00 26.64&	 6.89 &	 6.00 & 4.376&	 3.906	& 3.789\\
  HD~204613   &21 27 42.96& +57 19 18.86&	 8.86	& 8.22&  7.100&	 6.824&	 6.788\\
  
\hline
\end{tabular}
\end{table*}

\section{Radial velocity}
Radial velocities of the  programme stars are calculated using  a selected 
set of clean  unblended  lines in the spectra. Estimated mean radial velocities  
along with the standard deviation of the mean values are presented in 
Table 2. The literature values  are also presented  for a comparison. 
Reports on radial velocity variations for a large number of CH and 
barium stars are available
in literature (McClure (1984, 1997) and  McClure and Woodsworth (1990)).
 McClure (1997) has reported the radial velocity variations and orbital 
parameters for two sub-giant CH stars HD~122202 and HD~204613. These 
objects are confirmed  binaries. Although radial velocity variations are 
noticed in HD~55496, it is not  confirmed as binary. 
Our estimate also shows a difference of 7 Km$s^{-1}$ from the literature value.
Mild radial velocity
variations  are also noticed in HD~92545 and 
HD~107574 (North et al. 1992).
 Our radial velocity estimates of HD~104979 and HD~164922  show a difference 
of ${\sim}$ 15 Km$s^{-1}$ from the literature values. 

{\footnotesize
\begin{table*}
{Table 2: Radial velocities}\\
\begin{tabular}{cccc}
\hline
Star Name   & $ V_r$ km s$^{-1}$   &  $ V_{r}$ km s$^{-1}$&Reference  \\
            & our estimates       & from literature &   \\
\hline
  HD~55496    & 315.28 ${\pm}$ 0.80 & 322.00& 1 \\
  HD~89668    &22.84 ${\pm}$ 0.70 &23.0&2\\
  HD~92545    &$-$17.51 ${\pm}$ 0.65 & $-$16.65&  3 \\
  HD~104979   &$-$45.40 ${\pm}$ 0.42 & $-$29.62&4   \\
  HD~107574   &$-$16.33	${\pm}$ 0.72 &$-$29.40&1       \\
  HD~111721   &20.59 ${\pm}$ 0.67 & 21.40&1 \\
  HD~122202   &$-$7.40 ${\pm}$ 0.97 & $-$10.5& 6\\
  HD~126681   &$-$45.36 ${\pm}$ 0.46 &$-$45.58& 7\\
  HD~148897   &17.55 ${\pm}$ 0.71 & 18.40& 1 \\
  HD~164922   &34.86 ${\pm}$ 0.91 &20.29 & 8  \\
  HD~167768   &1.39 ${\pm}$ 0.42   &1.60& 1   \\
  HD~204613   &$-$89.53 ${\pm}$ 0.33 & $-$90.96&  5\\
  
\hline
\end{tabular}

1. Goncharov (2006), 2. Soubiran et al. (2008), 3. Siebert et al. (2011), 4. Massarotti et al. (2008), 5. Pourbaix et al. (2004),\\ 6. Luck and Bond (1991), 7. Santos et al. (2011), 8. Nidever et al. (2002)
\end{table*}
}
\section{ Temperatures from photometric data}

 Temperatures from photometric data are estimated following the procedure
 discussed in paper I. Here we mention a few points relevant to the present
work.   Colour-temperature calibrations of Alonso et al. (1996) are
used for photometric temperature determination. These calibrations  were 
derived by using a large number of lower main-sequence stars and sub-giants, 
whose temperatures were measured by the  infrared flux method, and hold 
within temperature and metallicity ranges of 
4000 K $\leq$ $T_{eff}$ $\leq$ 7000 K and metallicity between $-$2.5 and 0.0.
The  uncertainty in the temperature calibrations  is $\sim 100$\,K. 
Although the difference between 2MASS infrared photometric system and 
photometry data measured on the TCS(Telescopio Carlos Sanchez) system used by Alonso et al. to derive  
the $T_{eff}$ scales is very small, we have used the  
 necessary transformations between the different  photometric systems 
from  Ramirez and Melendez (2004) and Alonso et al. (1996, 1999). 
The  equations  are: \\

$J_{TCS}$ = $J_{2MASS}+0.001-0.049(J_{2MASS}-K_{2MASS})$\\
\\
$H_{TCS}$ = $H_{2MASS}-0.018+0.003(J_{2MASS}-K_{2MASS})$\\
\\
$K_{TCS}$ = $K_{2MASS}-0.014+0.034(J_{2MASS}-K_{2MASS})$\\
\\
$K_{J}$ = $K_{TCS}+0.042-0.019(((J_{TCS}-K_{TCS})-0.008)/0.910)$\\
\\
$(V-K)_{TCS}$ =$ 0.050+0.993(V-K_{J})$\\
\\
$\theta_{JK}$ =$ 0.582+0.799(J_{TCS}-K_{TCS})+0.085(J_{TCS}-K_{TCS})(J_{TCS}-K_{TCS})$\\
\\
$\theta_{JH}$ =$ 0.587+0.922(J_{TCS}-H_{TCS})+0.218(J_{TCS}-H_{TCS})(J_{TCS}-HT_{TCS})+0.016(M)(J_{TCS}-H_{TCS})$\\
\\
$\theta_{VK}$ =$ 0.555 + 0.195(V-K)_{TCS}+0.013(V-K)_{TCS}(V-K)_{TCS} - 0.008(V-K)_{TCS}(M)+0.009(M) - 0.002M^2$\\
\\
$T_{EFF(xy)}$ = $5040/\theta_{(xy)}$\\
\\
where M is the metallicity of the star, xy indicates the JK, JH and VK.
For two objects temperatures derived from both spectroscopic method and 
photometric method are similar. Among the rest for
 most of the objects the derived $T_{\rm eff}$ from V-K is ${\sim}$ 
350 K, and from J-H  is ${\sim}$ 300 K less than the adopted spectroscopic
 $T_{\rm eff}$.   The temperature calibrations  from the
 $T_{\rm eff}$ - $(J-H)$ and $T_{\rm eff}$ - $(V-K)$ relations  involve a 
metallicity ([Fe/H]) term. Estimates of  $T_{\rm eff}$  
at  four assumed  metallicity values (shown in parenthesis) are listed in 
Table 3.

{\footnotesize
\begin{table*}\tiny
{Table 3: Temperatures from  photometry }\\
\begin{tabular}{llllllllllll}
\hline
Star Name  &  $T_{eff}$  &  $T_{eff}(-0.5)$  &  $T_{eff}(-0.5)$  & $T_{eff}(-1.0)$  & $T_{eff}(-1.0)$  & T$_{eff}(-1.5)$  &
T$_{eff}(-1.5)$&$T_{eff}(-0.5)$  &  $T_{eff}(-1.0)$  & $T_{eff}(-1.5)$  & Spectroscopic   \\
           &  (J-K)      &   (J-H)          &   (V-K)          &  (J-H)          &    (V-K) &  (J-H) & (V-K) &(B-V) & (B-V) & (B-V) &estimates  \\
\hline
HD~55496  &     4542.90  &   4441.94  &    4502.61 &  4458.65 &  4487.14 &  4475.49 &  4475.76& 4774.22  & 4668.96  & 4591.12 &  4850 \\
HD~89668  &     4462.68  &   4502.02  &  4518.76   &  4535.63 &  4318.91 & 4302.05  & 4288.96 &4339.62  & 4246.01  & 4175.29  & 5400  \\  
HD~92545  &     6343.00  &   6422.81  &   6088.75  &  6436.34 &  6095.01 &  6449.92 &  6108.68 & 6159.33 &  6014.42 &  5914.10 & 6380      \\
HD~104979 &     5025.77  &   5056.49  &   4896.90  &  5073.13 &  4885.21 &  5089.87 &  4878.30& 4604.59  & 4503.92  & 4428.86  & 5060    \\
HD~107574 &     6605.48  &   6892.23  &   6774.10  &  6903.73 &  6795.30 &  6915.27 &  6825.87 & 6847.21  & 6681.19 &  6569.80& 6250   \\
HD~111721 &           -  &   4601.48  &    6451.12 &  4618.25 &  6464.85 &  4635.14 &  6486.97 &5011.10  & 4899.33  & 4817.63 & 5212    \\
HD~122202 &     6417.84  &   6878.47  &   6890.04  &  6901.64 &  6370.16 &  6382.14 &  6402.28 & 6329.12 &  6179.09 &  6076.03&6430 \\
HD~126681 &     5539.93  &   5508.00  &   5524.06  &  5540.22 &  5452.94 &  5448.38 &  5449.71 & 5629.53 &  5500.24  & 5408.49&5760   \\
HD~148897 &     3635.13  &   3885.78  &   3762.86  &  3901.90 &  3743.08 &  3918.15 &  3726.27 &  4015.00 &  3929.84 &  3864.45 & 4285   \\
HD~164922 &     5412.07  &   5422.44  &   5191.94  &  5438.65 &  5183.77 &  5454.94 & 5180.94  &  5038.84 &  4926.30 &  4844.15& 5400\\
HD~167768 &     4841.08  &   4060.94  &   4910.98  &  4077.32 &  4899.44 &  4093.83 &  4892.70 & 4799.45  & 4693.50  & 4615.26 & 5070  \\
HD~204613 &     6070.18  &   5869.04  &    5841.82 &  5884.32 &  5843.53 &  5899.68 &  5852.03 & 5494.19 &  5368.80  & 5279.24 &5875  \\
\hline
\end{tabular}

The numbers in the parenthesis indicate the metallicity values at which the temperatures are calculated. Temperatures are given in Kelvin\\
\end{table*}
}

\section{Stellar Atmospheric parameters}  

 The  set of  Fe I and Fe II lines used for the present analysis 
to find the stellar atmospheric parameters are listed in Tables 4A and 4B.
The  excitation potential of the lines are  in the range  
0.0 - 5.0 eV and equivalent width in the range 20 {\rm \AA} to 180 {\rm \AA}.
 We have assumed local thermodynamic equilibrium (LTE) for our calculations.
 A recent version of MOOG of Sneden (1973) is used.
Model atmospheres  (available at http: //cfaku5.cfa.harvard.edu/  and labelled 
with a suffix odfnew) were selected from 
the Kurucz grid of model atmospheres with no convective over shooting. 
 Solar abundances are taken from Asplund et al.(2005). 

The microturbulent velocity is estimated at a given effective temperature 
by demanding that there be no dependence of the derived Fe I abundance 
on the equivalent width of the corresponding lines. 

The effective temperature is determined by making the slope of the 
abundance versus excitation potential of Fe I lines to be nearly zero. 
The initial value of temperature is taken from the photometric estimates
and  arrived at a final value by an iterative method with the slope 
 nearly equal to zero. Figures 3 and 4 show abundances of Fe I and Fe II 
as a function of excitation potential and equivalent width respectively. 

The surface gravity is fixed at a value that gives same abundances for
Fe I and Fe II lines. 
Derived atmospheric parameters are listed in Table 5.
{\footnotesize
\begin{table*}
{\bf  Table 4A: Fe lines used for deriving atmospheric parameters for the first 6 objects}\\
\begin{tabular}{lllcllllll}
\hline
Wavelength({\rm \AA})&    Element    &    E$_{low}$(ev) &   log gf&  HD~55496 &HD~89668&HD~92545&HD~104979&HD~107574&HD~111721\\

\hline

 4062.440 &   Fe I  &   2.850&-0.860 &    -    &  -     & 110.3  &      -&  -&- \\     
 4114.440 &   &   2.830 &   -1.300    &     -   & -	&       &    -&-&- \\
 4132.900 &   &   2.850 &   -1.010    &     -	 &     & -	 &    -   &  -&- \\
 4143.870 &   &   1.560 &   -0.510	&     -  &     & -	 &    -    & -&- \\
 4147.670 &   &   1.490 &   -2.100	&     -	 &     & -	 &    -    & -&- \\
 4153.900 &   &   3.400 &   -0.320	 &    -	 &     & -	 &    -    & -&- \\
 4154.500 &   &   2.830 &   -0.690	 &    -	 &     & -	 &    -     &-&- \\
 4184.890 &   &   2.830 &   -0.870	&     -	 &     & 151.2   &   143.1 &-&- \\
\hline
\end{tabular}

This table is available in its entirety in online only. A portion is shown here for guidance regarding its form and content.\\ 
\end{table*}
}

{\footnotesize
\begin{table*}
{\bf  Table 4B: Fe lines used for deriving atmospheric parameters for the next 6 objects}\\
\begin{tabular}{lllcllllll}
\hline
Wavelength({\rm \AA})&    Element    &    E$_{low}$(ev) &   log gf&  HD~122202 &HD~126681&HD~148897&HD~164922&HD~167768&HD~204613\\

\hline
4062.440 &   Fe I  &   2.850&-0.860 &- & - &  - &  - &  - &-  102.9 \\
 4114.440 &   &   2.830 &   -1.300  & - & - & 129.8 &  - &- & -\\
 4132.900 &   &   2.850 &   -1.010  & - & - & - &   - &- & 107.6 \\
 4143.870 &   &   1.560 &   -0.51   & - & - &  - &  - &- & -\\
 4147.670 &   &   1.490 &   -2.100  & - & - &  - &  - &- & 114.6\\
 4153.900 &   &   3.400 &   -0.320  & - & - & - &   - &- &- \\
 4154.500 &   &   2.830 &   -0.690  & - & - &  - &  - &- & -\\
 4184.890 &   &   2.830 &   -0.870  &- &- & - & - &- &  105.0\\

\hline
\end{tabular}

This table is available in its entirety in online only. A portion is shown here for guidance regarding its form and content.\\
\end{table*}
}
\begin{figure}
\centering
\includegraphics[height=8cm,width=8cm]{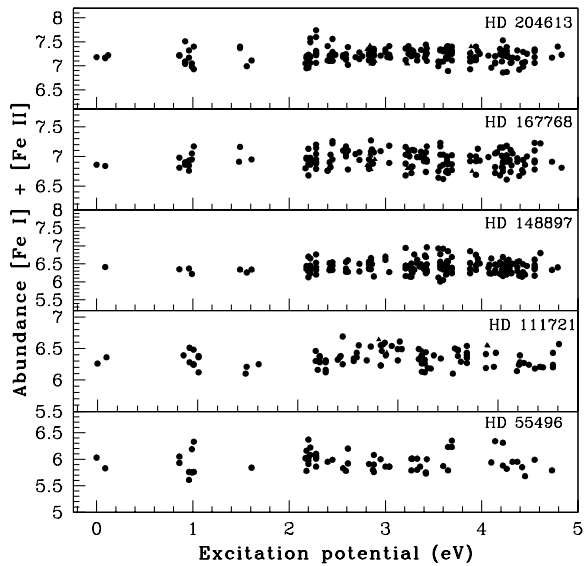}
\caption{The iron abundances of stars are shown for
individual Fe I and Fe II lines as a function of excitation potential.
The solid circles indicate Fe I lines and solid triangles  indicate 
Fe II lines.}
\end{figure}

\begin{figure}
\centering
\includegraphics[height=8cm,width=8cm]{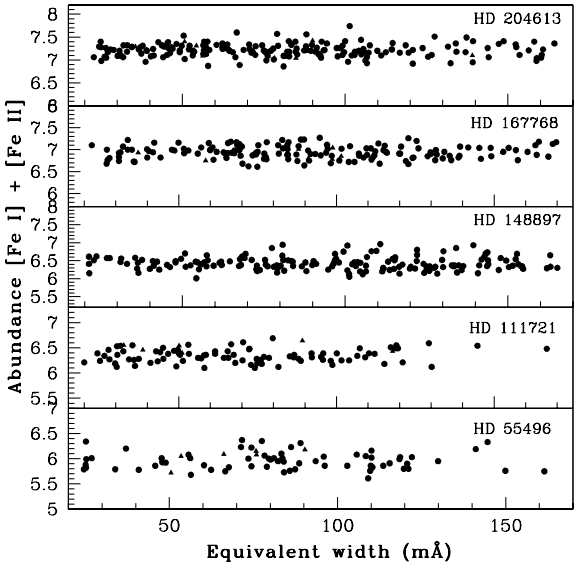}
\caption{The iron abundances of stars  are shown for individual Fe I 
and Fe II lines as a function of equivalent width. The solid circles 
indicate Fe I lines and solid triangles indicate Fe II lines.}
\end{figure}

 \begin{table*}
{\bf  Table 5: Derived atmospheric parameters and Carbon isotopic ratios for the programme stars}\\
\begin{tabular}{lcccccc}
\hline
Star Name.   & $T_{eff}$  &  log g  &  $\zeta $ &[Fe I/H] &[Fe II/H]&$C^{12}/C^{13}$    \\
             &   K        &         &  km s$^{-1}$ &      &             \\
\hline
  HD~55496    &4850&2.05  &1.52  &-1.49  &-1.41&4   \\
  HD~89668    &5400&4.35&2.35&-0.13&-0.19&19.1\\
  HD~92545    &6380&4.65  &1.45 &-0.21  & -0.22 & -\\
  HD~104979   &5060&2.67  & 1.55 &-0.26  &-0.31 &9.9   \\
  HD~107574   &6250	& 2.9 &1.35  & -0.65 &  -0.60 &- \\
  HD~111721   &5212&2.6  &1.30  &-1.11  &-1.11&  - \\
  HD~122202   &6430&4.0&2.08&-0.63&-0.65&13.2\\
  HD~126681   &5760&4.65&0.9&-0.90&-0.92&-\\
  HD~148897   &4285&0.6 &1.83  &-1.02  &-0.99& 13  \\
  HD~164922   &5400&4.3 &0.09  &0.22  &0.23& 12\\
  HD~167768   &5070&2.55  &1.49  &-0.51  &-0.56&-   \\
  HD~204613   &5875  &4.2  &1.22  &-0.24  &-0.24&11.1   \\
  
\hline
\end{tabular}
\end{table*}
\section{Abundance analysis}
Elemental abundances are  calculated  from  the measured  
equivalent widths of lines due to neutral and ionized elements using a recent 
version of MOOG of Sneden (1973) and  the adopted model atmospheres. 
A master line list of all the elements is generated comparing the 
spectra of the programme stars with the spectrum of Arcturus. The 
presented line lists contain only those lines which are used for  
abundance calculation. Even though we could detect many lines for
each element, only a  few were usable for abundance calculation, the others
being either distorted or blended with contributions from other species.
The log gf values of the atomic lines are taken from literature consulting
various sources, such as,
 Aoki et al. (2005, 2007), Goswami et al. (2006), 
Jonsell et al. (2006), Luck and Bond (1991), Sneden et al. (1996), 
and  Kurucz  atomic line database  (Kurucz 1995a,b).
The log gf values for a  few La lines are taken from
Lawler et al. (2001). 
We have estimated  abundances for  many elements  Na, Mg, Ca, Sc, 
Ti, V, Cr, Mn, Co,  Ni, Zn and for heavy elements Sr, Y, Zr, Ba, La, Ce, Pr, 
Nd, Sm, Eu and Dy. 
For the elements Sc, V, Mn, Ba, La and Eu, spectrum synthesis
 is used to find the abundances considering  hyperfine 
structure. 
The line lists for each region that is synthesised are taken from Kurucz 
atomic line list (http://www.cfa.harvard.edu/amdata/ampdata/kurucz23/\\
sekur.html). A few examples of spectrum synthesis calculations 
 are shown  in  Figures  5, 6 and 7.

 Derived abundance ratios with respect to iron are listed in Table 6. 
In Table 7, we have presented [ls/Fe], [hs/Fe] and [hs/ls] values,
where ls represents light s-process elements Sr, Y and Zr and hs
represents heavy s-process elements Ba, La, Ce, Nd and Sm. Lines used 
for the abundance calculation of these elements are listed in 
Tables 8A, 8B, 9A and 9B. 
\begin{figure}
\centering
\includegraphics[height=8cm,width=8cm]{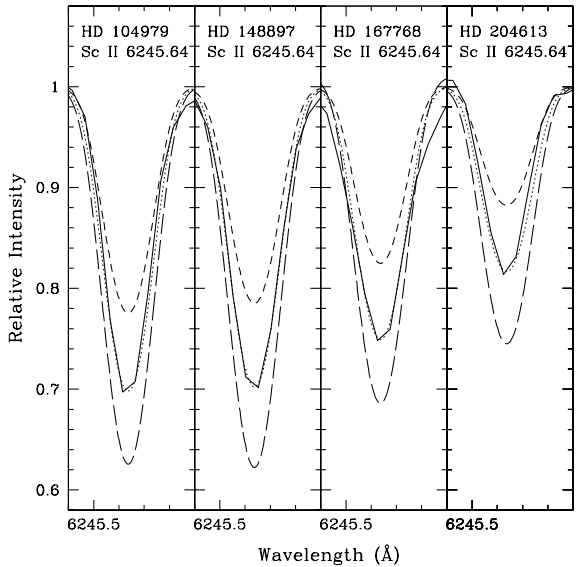}
\caption{ Spectral-synthesis fits of Sc II line at 6245.64 {\rm \AA}. The 
dotted lines indicate the synthesized spectra and the solid
lines indicate the observed line profiles. Two alternative synthetic spectra
for [X/Fe] = +0.3 (long-dashed line) and [X/Fe] = -0.3 (short-dashed
line) are shown to demonstrate the sensitivity of the line strength to the
abundances.}
\end{figure}
%label(fig 5)

\begin{figure}
\centering
\includegraphics[height=8cm,width=8cm]{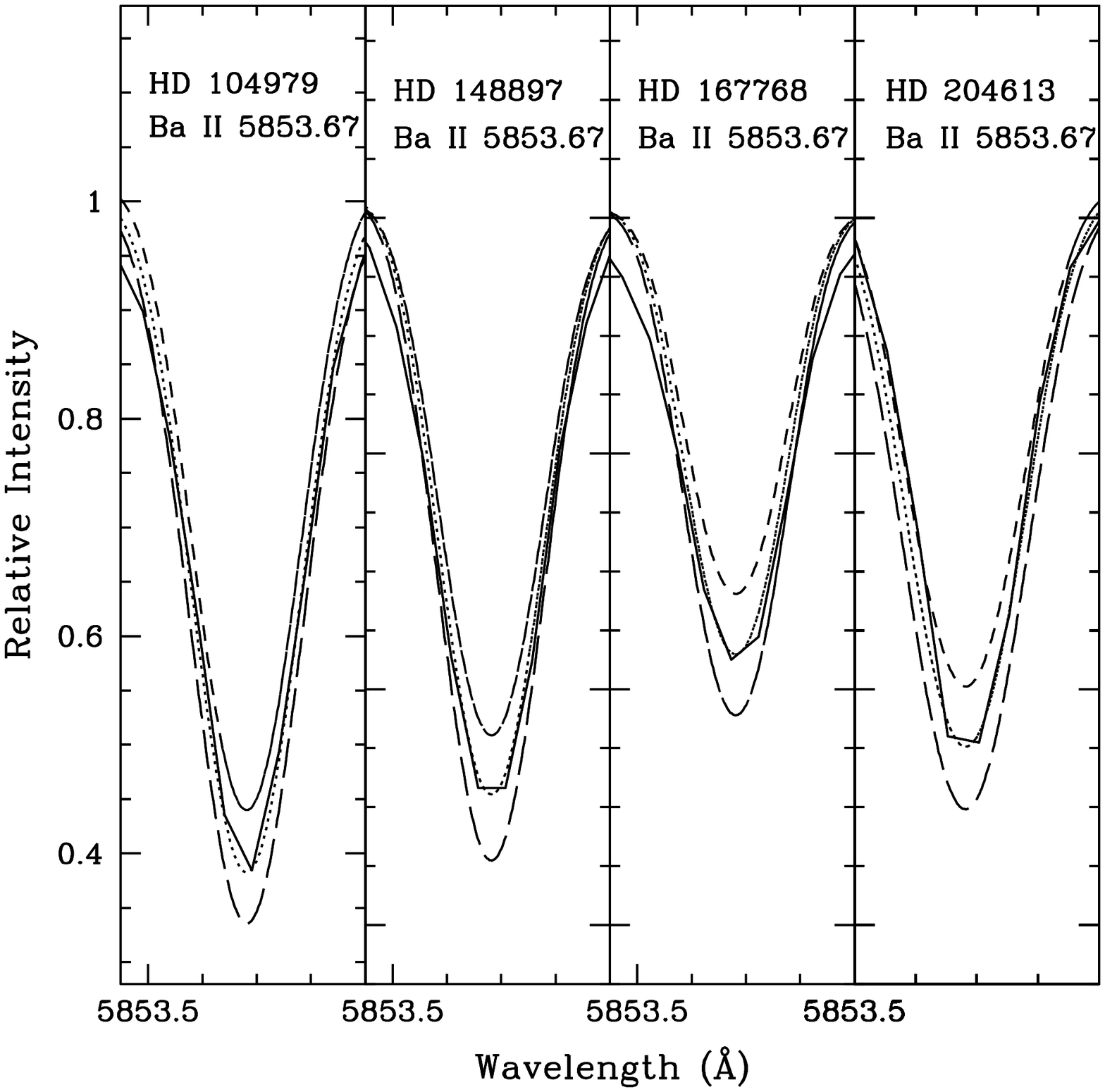}
\caption{ Spectral-synthesis fits of Ba II line at 5853.67 {\rm \AA}. 
The dotted lines indicate the synthesized spectra and the solid
lines indicate the observed line profiles. Two alternative synthetic spectra
for [X/Fe] = +0.3 (long-dashed line) and [X/Fe] = -0.3 (short-dashed
line) are shown to demonstrate the sensitivity of the line strength to the
abundances.}
\end{figure}
%\label(fig 6)

\begin{figure}
\centering
\includegraphics[height=8cm,width=8cm]{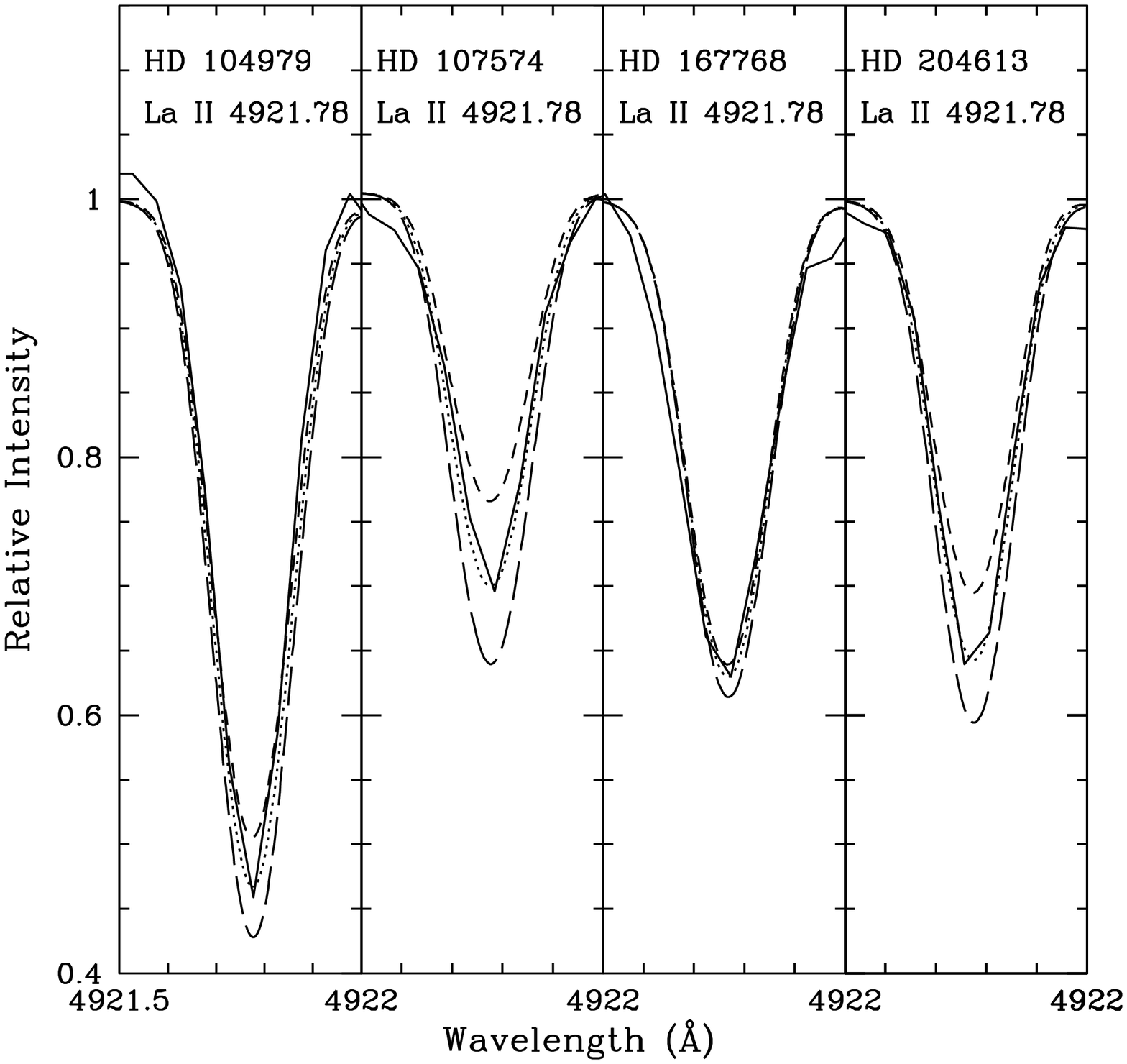}
\caption{ Spectral-synthesis fits of La II line at 4921.78 {\rm \AA}. 
The dotted lines indicate the synthesized spectra and the solid
lines indicate the observed line profiles. Two alternative synthetic spectra
for [X/Fe] = +0.3 (long-dashed line) and [X/Fe] = -0.3 (short-dashed
line) are shown to demonstrate the sensitivity of the line strength to the
abundances.}
\end{figure}

\subsection{Carbon}
We have derived the Carbon abundance for our objects whenever possible 
using the spectrum synthesis calculation of C I line at 5380.337 \AA.
The line list is generated  from Kurucz atomic and molecular line database(http://www.cfa.harvard.edu/amdata/ampdata/\\
kurucz23/sekur.html).
This line appears heavily distorted  in the spectra of stars        
HD 89668, HD 111721 and HD~126681 and hence C abundance  could not be 
determined for these objects from this line.
In case of  HD~148897, a very  week feature of C I at 5380.337 $\AA$ is detected; 
however, this line could not be used  for abundance determination 
using the spectrum synthesis calculation.
For  the stars HD~148897 and HD~111721
 the Carbon abundance is determined using spectrum synthesis calculation
of the  CH band at 4300 $\AA$.
 For HD~126681 we could not find C abundance due to severe line distortion and
blending throughout the spectrum.  Estimated  [C/Fe] ratios  are
 listed in Table 6.  We have derived the  $^{12}$C/$^{13}$C
ratio for seven objects from spectrum  synthesis of the  CH band.
The initial $^{12}$C/$^{13}$C is fixed at solar value and then varied  to fit
the observed spectrum for
the determined Carbon abundances. The estimated values lie in the range 4 to 19 
and
are presented  in Table 5 along with the atmospheric parameters.  The line 
list for the synthesis of CH band is taken from the Kurucz database for
molecular lines. We have derived a [C/H] value of -0.23 and -1.23 for two
cyanogen weak giants HD~104979 and HD~148897. For these objects Luck (1991)
reported the [C/H] values  $-$0.38 and $-$0.94 respectively. We have
determined a Carbon abundance of 8.68 dex for HD~204613  while Smith
et al. (1993) reported  8.91 dex for the same object. North et al. (1994)
has given the [C/H] ratios  $-$0.07 and $-$0.03 respectively for HD~92545
and HD~107574. We have derived  slightly lower [C/H] values for these objects.
For HD~92545 and HD~107574 our estimated  [C/H] values are  $-$0.37 and
$-$0.18 respectively.  Masseron et al. (2010) listed the [C/Fe] ratio of
these two objects as 0.32 and 0.39 respectively. Carbon abundance for
HD~122202 is not available in the literature. We have derived
 $-$0.13 for [C/H] and 0.5 for [C/Fe]  in  this object. Our  estimate of
[C/H] ${\sim}$ -0.48 and  [C/Fe] ${\sim}$ 0.03 for  HD~167786 are  in good
agreement with the estimate  of -0.63 and  $-$0.02 respectively of
Luck and Heiter (2007).

\subsection{Na and Al}
 The abundance of sodium is derived for all the objects except HD~122202. For
most of the objects we have used the lines at 5682.65 and 5688.22 {\rm \AA}. 
We have also used the doublet lines at 5890.9 and 5895.9 {\rm \AA} for 
 determination of sodium abundances. 
However the resonance lines are sensitive to non-LTE effects. 
The observed LTE abundance ranges between $-$0.29 to 0.49 in the programme stars. 

Even though we could measure a few Al lines in our programme stars spectra, these
 are not usable for abundance determination of Al. 

\subsection{ Mg, Si, Ca, Sc, Ti, V}
We have measured several lines due to these elements. Except for 
HD~92545 and HD~104979 that show near-solar values, all other stars  
show  mild enhancement of Mg with [Mg/Fe] $\geq$ 0.15. [Mg/Fe] in HD~148897 
with a metallicity of $-$1.02 is ${\sim}$ 0.63, slightly higher than as 
expected for classical enhancement of ${\alpha}$-elements in stars with 
[Fe/H] ${\sim}$ $-$1.0 (Goswami \& Prantzos 2000).  
Abundance of Si could not be estimated as none of the Si lines are found 
usable for abundance determination. Ca shows a near solar value in HD~92545, 
HD~104979, HD~126681 and HD~164922. In the rest of the  objects Ca is
found to be  mildly  enhanced.  

 Sc abundance is determined using spectrum synthesis calculation of 
Sc II line at 6245.63 {\rm \AA} considering hyperfine structure from 
Prochaska and Mc William (2000). We could determine Sc abundance in 
seven of the programme stars. Except HD~204613 with [Sc/Fe] value 0.17 
all the other objects show mild underabundance of Sc.  

Mild overabundance or near-solar abundance for Ti is noticed in all the 
programme stars except for HD~89668 and HD~92545. More than ten good lines 
of Ti are used for abundance determination.

Abundance of V is estimated from  spectrum synthesis calculation of V I line 
at 5727.028 ${\rm \AA}$ taking into account the hyperfine components
from Kurucz database. We could determine V abundance only in six objects. 
While HD~148897 shows a mild under abundance  with [V/Fe] $\approx$ -0.18, 
HD~89668, HD~104979, HD~167768, and HD~204613 show near-solar values. 
HD~164922 shows a mild overabundance with [V/Fe] = 0.40. 
We have detected more than   16 V I lines but  only one or two are 
usable   for the determination of  abundance; other lines 
appear either blended or distorted in the spectra. 

\subsection{Cr, Co, Mn, Ni, Zn}
 HD~122202, HD~107574, HD~148897 and HD~164922 show a near-solar abundance 
for Cr.
The rest of the stars in our sample are mildly underabundant 
in Cr.  HD~55496 however shows a larger  underabundance with 
[Cr/Fe] = $-$0.35. Cr abundances measured using Cr II lines 
whenever possible also show similar trends.

 Mn abundance is obtained using spectrum synthesis calculation of 
6013.51 {\rm \AA} line taking into account the  hyperfine structures  from 
Prochaska \& McWilliam (2000).
Except for HD~89668 and HD~164922, that show a mild overabundance with 
[Mn/Fe] ${\sim}$ 0.34 and 0.14 respectively,
 the rest of the objects  show  underabundance  with [Mn/Fe] $\le$ $-$0.23. 

 Except HD~92545 with [Co/Fe] ${\sim}$ 0.80, all other
stars in our sample show near-solar values or mild underabundance for Co. 

 Abundances of Ni measured from Ni I lines give
near-solar values for all the stars. 

  HD~122202 is  mildly overabundant in Zn  with [Zn/Fe] ${\sim}$ 0.59.  
The rest of the objects show near-solar values.

\subsection{ Sr, Y, Zr}
The abundance of Sr  is estimated in seven stars  using  Sr I line at 
4607.327 {\rm \AA}.  Sr is overabundant in HD~89668 and HD~204613 
 with  [Sr/Fe] $>$ 1.0. The other five objects HD~55496, HD~104979, 
HD~148897, HD~164922 and HD~167768 give [Sr/Fe] ratios in the range
 0.30 and 0.99. 
 Abundance of  Sr  could not be estimated in the remaining objects as the  
line at  4607.327 {\rm \AA} appears distorted in their spectra. None of 
the Sr II lines detected  are found suitable for abundance estimate of Sr.

The abundance of Y is measured in all the stars. Y is overabundant in 
HD~122202 and HD~107574  with [Y/Fe] ratio $\ge$ 1.0. 
HD~204613 and  HD~55496 show [Y/Fe] values of 0.97 and 0.85 respectively. 
The remaining stars show near-solar values or mild overabundance. 

We could derive Zr abundance for five stars.  HD~204613 and 
HD~104979 show overabundance 
with [Zr/Fe] values  1.14 and 0.85 respectively.  The rest show mild 
enhancement with [Zr/Fe] $\ge$ 0.2. 

\subsection{ Ba, La, Ce, Pr, Nd, Sm, Eu, Dy}

 As many lines due to Ce, Pr, Nd, Sm and Dy could be measured on our 
spectra the standard abundance determination method using equivalent width 
measurements 
are used for  abundance estimates. Spectrum synthesis 
calculation is also performed  for Ba, La and Eu.   We have estimated the 
abundance for  Ba and Ce in all the stars.  

 Barium (Ba): We have determined Ba abundance from  spectrum  synthesis  
calculation using Ba II line at 5853.668 ${\rm \AA} $ considering hyperfine 
components from Mcwilliam (1998). Four stars in our sample  HD~92545, 
HD~104979, HD~107574 and HD~204613 show overabundance 
with [Ba/Fe]  $\geq$ 0.9. HD~122202, HD~55496 and HD~126681 show only a mild 
overabundance. Four objects HD~89668, HD~111721, HD~148897, and HD~167768
show  underabundance with [Ba/Fe]  in the range $-$0.09 to $-$0.65 (Table 11).

Lanthanum (La): We have derived La abundance for all the programme stars 
except for HD~55496 and HD~126681 from  spectrum 
synthesis calculation of La II line at 4921.77 ${\rm \AA}$ 
considering  hyperfine components from Jonsell et al. (2006). 
Except for HD~111721, HD~148897, HD~164922 and HD~167768, La in  all other stars are 
found to be  
overabundant  with [La/Fe] $\ge$ 0.9. HD~111721, HD~148897, HD~164922 and HD~167768 
show  [La/Fe] of 0.31, 0.29, 0.15 and $-$0.54 respectively.

Cerium (Ce): We have derived Ce abundance for all the programme stars. 
Six of the programme stars, HD~89668, HD~92545, HD~104979, HD~111721, HD~122202 
and HD~204613 show overabundance with [Ce/Fe] $\ge$ 1.0. Estimated [Ce/Fe]
for HD~107574 is ${\sim}$ 0.6.
 While two stars 
HD~55496 and HD~167768  show almost near-solar values for [Ce/Fe], 
HD~148897 and HD~164922 show mild underabundance with [Ce/Fe] $\approx$ $-$0.10. 

Praseodymium (Pr):  We could derive Pr abundance for five programme stars 
mainly using  the Pr II line at 5292.619 {\rm \AA}.
A mild over enhancement of Pr is seen in  HD~55496 with
[Pr/Fe] ${\sim}$ 0.43; the rest show overabundance with [Pr/Fe] $\ge$ 1.0.

Neodymium (Nd): Abundance of  Nd is estimated for seven programme stars. 
 Two stars HD~148897 and HD~167768 give 
[Nd/Fe] values  ${\sim}$ 0.13 and 0.65 respectively. HD~111721 shows a 
large overabundance with 
[Nd/Fe] ${\sim}$ 2.1. All other stars show  overabundance with 
[Nd/Fe] $\ge$ 1.0. 

Samarium (Sm): HD~148897 shows a mild overabundance
with [Sm/Fe] ${\sim}$ 0.58. HD~89668, HD~104979, HD~122202, HD~126681 and 
HD~204613 show overabundance with [Sm/Fe] $\ge$ 1.0. Estimated [Sm/Fe] in 
HD~167768 is ${\sim}$ 0.90.

Europium (Eu): We could determine Eu abundance in four of the programme stars 
using  spectrum synthesis of Eu II lines at 6645.130 {\rm \AA} by considering  the hyperfine components from 
Worley et al. (2013). Eu shows mild overabundance in HD~89668, HD~92545 and 
HD~167768 with [Eu/Fe] ${\sim}$ 0.38, 0.40 and 0.26 respectively. 
 HD~ 204613 shows a near-solar value with [Eu/Fe] ${\sim}$ 0.06.

Dysprosium (Dy): We could derive Dy abundance for three objects HD~148897, 
HD~167768 and HD~204613 using  Dy II lines at  4103.310 {\rm \AA} and 
4923.167 {\rm \AA}.  HD~167768 and HD~204613 show overabundance with 
[Dy/Fe] $\ge$ 1.0. HD~148897 shows a near-solar value of ${\sim}$ 0.02.

{\footnotesize
\begin{table*}\tiny
{\bf  Table 6: Elemental abundances}
\begin{tabular}{lllllllllllllll}
\hline
Star Name&[C/Fe]&[Na I/Fe]&[Mg I/Fe]&[Ca I/Fe]&[Sc II/Fe]&[Ti I/Fe]&[Ti II/Fe]&[V I/Fe]&[Cr I/Fe]&[Cr II/Fe]&[Mn I/Fe]&[Co I/Fe]&[Ni I/Fe]&[Zn I/Fe] \\
\hline
Sub giant\\
CH stars\\
HD~122202   &  0.50&    -   &  0.32 & 0.33 & -   &  -   &  0.36 & -   &  0.11  &-   &  -  &   -   &  0.14 &  0.59\\
HD~204613   &      0.49&0.08 & 0.11 & 0.13 & 0.17 & 0.26 & 0.41 & -0.01& -0.06 &0.08 & -0.33 &-0.19& 0.04 &  -\\
$\#$CH stars\\
$^*$HD 55496&  1.01     & 0.4 &  0.33&  0.46&  - &    -0.1 & -0.16 &0.19 & -0.35 &-0.21 &-   &  -   &  -0.18 & 0.02\\
HD~89668    &   0.05   &0.08 & 0.28 & 0.63&  0.0  & -0.26& -0.38& -   &  -0.09 &-0.27 &0.34 & -0.23 &-0.12\\
HD~92545    &   0.68   &0.01 & -0.09& -0.05 &-  &   0.03 & 0.6 &  -   &  -0.15& -  &   -   &  0.8  & 0.01 &-\\
HD~104979  &  0.03     &-0.01& 0.07 & 0.05 & -0.01& 0.14 & 0.32 & 0.05 & -0.02& 0.06  &-0.23& 0.28 & 0.04  & -0.03\\
HD~107574  &   0.47    &0.49 & -    & 0.19 & -0.13 &0.32 & 0.34&  -   &  0.1 &  -0.26& -   &  -  &   0.01  & -\\
HD~111721  &  0.08   &  0.07 & 0.46 & 0.41 & -  &   0.46 & 0.03 & -   &  -0.21& -0.31& -  &   -   &  -0.07 & -\\
HD~126681 &   -    & -0.26 &0.44 & 0.07 & -  &   0.52 & 0.60&  -  &   0.1 &  -   &  -   &  -    & -0.08 & -\\
HD~148897 &   -0.21     &-0.29& 0.63&  0.19 & -0.33 &0.11 & 0.46 & -0.18& -0.22& 0.0  & -0.54& 0.06  &-0.13 & -0.26\\
HD~164922&   -0.15      &-0.02&0.36& -0.07&  -0.47 &0.25& 0.05 & 0.40&0.04& 0.06 & 0.14  &0.13&0.06 &0.19 \\
HD~167768 & 0.03       &0.04 & 0.17 & 0.22 & -0.04 &0.17 & 0.41 & 0.2 &  -0.09 &-0.21& -0.56& -0.03 &-0.09 & 0.18\\

\hline
\end{tabular}
$^\#$ Objects from the CH star catalogue of Bartkevicius (1996)\\
$^*$ Objects are also included in Ba star catalogue of L\"u (1991\\
\end{table*}
}
{\footnotesize
\begin{table*}\tiny
{\bf  Table 6: continued}\\
\begin{tabular}{llllllllllll}
\hline
Star Name& [Sr I/Fe]&[Y II/Fe]&[Zr II/Fe]&[Ba II/Fe]&[La II/Fe]&[Ce II/Fe]&[Pr II/Fe]&[Nd II/Fe]&[Sm II/Fe]&[Eu II/Fe]&[Dy II/Fe]\\
\hline
Sub gian\\
CH stars\\
HD~122202 & -  &  1.44& -    & 0.33  &0.9  & 1.62 & 1.26 &-  &  1.77 &-  &  -\\
HD~204613 & 1.71& 0.97& 1.14&  1.04 & 1.21 & 1.24 & 1.52& 1.02& 1.61& 0.06 &1.77\\
$\#$CH stars\\
$^*$HD~ 55496& 0.82& 0.85& 0.52  &0.57 & -  &   0.13 & 0.43& -   & -   & -   & -\\
HD~89668 &  1.06& 0.55& -  &   -0.24 &1.87 & 1.52 & 1.66 &1.44& 1.23& 0.38 &-\\
HD~92545&   -    &0.23& -    & 0.91 & 0.95 & 1.6 &  -   & -   & -  &  -  &  -\\
HD~104979 & 0.99& 0.71& 0.85 & 0.94  &1.11 & 1.06  &1.04 &1.13 &1.17 &0.40 &-\\
HD~107574 & -   & 1.02& -    & 0.97 & 1.04 & 0.6  & -   & - &   -   & - &   -\\
HD~111721 & -   & 0.05 &-    & -0.09 &0.31&  1.6&   -   & 2.1 & -   & -  &  -\\
HD~126681&  -   & 0.02 &-    & 0.27 & -    & 0.67 & -   & 1.2 & 1.07 &-   & -\\
HD~148897&  0.31& 0.03 &-0.47& -0.65& 0.29  &-0.16& -   & 0.13 &0.58& -   & 0.02\\
HD~164922 &         0.79 & 0.14 &-&  0.28 &0.15&  -0.09&- &-&-& -& -\\
HD~167768&  0.77& 0.56& 0.2&   -0.36 &-0.54 &0.06 & -    &0.65 &0.9 & 0.26 &1.04\\

\hline
\end{tabular}
$^\#$ Objects from the CH star catalogue of Bartkevicius (1996)\\
$^*$ Objects are also included in Ba star catalogue of L\"u (1991\\
\end{table*}
}

{\footnotesize
\begin{table*}
{Table 7: Observed values for [Fe/H], [ls/Fe], [hs/Fe]  and [hs/ls]}\\
\begin{tabular}{lcccccc}
\hline
Star Name & [Fe/H]  & [ls/Fe] & [hs/Fe] & [hs/ls] & Remarks   \\
\hline
HD 55496  &-1.49 & 0.73 &0.38 &-0.35&1\\
HD 89668  &-0.13 & 0.81 &1.16 &0.35&1\\
HD 92545  &-0.21 & 0.23 &1.15& 0.92&1\\
HD 104979& -0.26 & 0.85 &1.03& 0.18&1\\
HD 104979&-0.47   & 0.6  &1.0 &0.4&2\\
HD 107574& -0.48 & 1.02 &0.87& -0.15&1\\
HD 111721& -1.11 & 0.05 &0.98 &0.93&1\\
HD 122202& -0.63 & 1.44 &1.16 &-0.28&1\\
HD 126681& -0.90 & 0.02 &0.80 &0.78&1\\
HD 148897& -1.02 &-0.13 &0.01& 0.14&1\\
HD 164922&0.22 & 0.47 & 0.10 & -0.37&1\\
HD 167768& -0.51 & 0.51 &0.14& -0.37&1\\
HD 204613& -0.24 & 1.27 &1.16& -0.11&1\\
HD 204613&-0.35 & 1.0& 0.6 & -0.4 &2\\
\hline
\end{tabular}

1. Our work; 2: Busso et al. (2001) \\
\end{table*}
}
{\footnotesize
\begin{table*}
{\bf  Table 8A: Equivalent widths in m{\rm \AA} of lines used for the calculation of light element abundances for first 6 objects}\\
\begin{tabular}{lllcllllll}
\hline
Wavelength({\rm \AA})&    Element    &    E$_{low}$(ev) &   log gf&  HD~55496 &HD~89668&HD~92545&HD~104979&HD~107574&HD~111721\\

\hline
5682.650  &   Na I  &  2.100 & -0.700 &                      58.07     &      208.7	           &    63.7                  &     115.9          &           50.1	  &  -	\\      	   
5688.220  &         &  2.100 & -0.400 &                     -  &      205.3	           &    89.9     	      &	   131.6 	 	  &   - &     29.9	 	 \\  	  
5889.950  &         &  0.000 &  0.100 &                      275.5     &	      - 	   &    -		      &	   450.2 	 	  &     253.8	  &     263.3	 	\\ 	   	  
5895.920  &         &  0.000 & -0.200 & -      &       -	   &    225.4 		      &	   379.5 	 	  &     226.1	  &     245.6	 	 \\	  
4702.990  &   Mg I  &  4.350 & -0.666 &    103.2     &	    -  &    172.8 		      &	   188.9 	 	  &   - &     143.1	 	 \\ 
6318.720   &         &  5.108 & -1.730 &          15.88  &  86.1&    82.6   &	   137.4 	 	  &   - &     79.4	 	  \\  	   	  
5528.000   &         &  4.350 & -0.490 &                      142.3     & -  &    162.6 &	   202.4  &   - &     139.0	 	 \\	  
4098.500   &   Ca I  &  2.525 & -0.540 &                    -      &       - 	   &    -		      &	 -	  &	      - &   -	 \\

\hline
\end{tabular}

This table is available in its entirety in online only. A portion is shown here for guidance regarding its form and content.\\ 
\end{table*}
}
{\footnotesize
\begin{table*}
{\bf  Table 8B: Equivalent widths in m{\rm \AA} of lines used for the calculation of light element abundances for next 6 objects}\\
\begin{tabular}{lllcllllll}
\hline
Wavelength({\rm \AA})&    Element    &    E$_{low}$(ev) &   log gf&HD~122202&HD~126681&HD~148897 & HD~164922& HD~167768 &HD~204613\\

\hline
5682.650  &   Na I  &  2.100 & -0.700 &        -        &   22.8     &	  73.7	    & 	   137.6       &  93.2		           &     87.5   \\  
5688.220  &         &  2.100 & -0.400 &        -        &   -       &     93.0	    & 	   143.4       &  118.7		 	  &       106.2 \\
5889.950  &         &  0.000 &  0.100 &        -        &   263.2    &	   396.5    & 	   -    &  379.1		 	  &       377.3 	\\ 5895.920  &         &  0.000 & -0.200 &        -        &   234.5    &	   340.2    & 	   601.4       &  329.7		 	  &       -    	 \\	  
4702.990  &   Mg I  &  4.350 & -0.666 &        147.3     &   -       &	   186.2    & 	   334.3       &  182.4		 	  &       232.6 	 \\ 
6318.720   &         &  5.108 & -1.730 &        -        &   60.7     &	   35.3     & 	   133.2       &  35.5		 	  &       88.3  \\
5528.000   &         &  4.350 & -0.490 &        -        &   156.7    &	   212.2    & 	   320	       &  205.3		 	  &       201.8 \\
4098.500   &   Ca I  &  2.525 & -0.540 &        -        &   -       &	 -   & 	   130.6       &  -		 	  &       79.2  	 \\

\hline
\end{tabular}

This table is available in its entirety in online only. A portion is shown here for guidance regarding its form and content.\\ 
\end{table*}
}

{\footnotesize
\begin{table*}
{\bf  Table 9A: Equivalent widths in m{\rm \AA} of lines used for  abundance determination  of  heavy elements for the first 6 objects}\\
\begin{tabular}{lllcllllll}
\hline
Wavelength({\rm \AA})&    Element    &    E$_{low}$(ev) &   log gf&  HD~55496&HD~89668&HD~92545&HD~104979&HD~107574&HD~111721\\
\hline
4607.327 & Sr I  & 0.000 &  -0.570 & 36.42 &  140.9 & -      &      91.55 & -     & -       \\   
4854.863 &Y II   & 0.992 &  -0.380 & -     &  90.5  & 57.6   &      120.6 & -     & -        \\
4883.685 &       & 1.084 &   0.070 & -     &  74.4  & 89.9   &      121.2 & 106.8 & 51.7     \\
5087.416 &       & 1.080 &  -0.170 & 103.5 &  -     & 72.1   &      91.82 & -     & 56.2     \\
5119.112 &       & 0.992 &  -1.360 & 41.16 &  -     & 27.4   &      55.28 & -     & -        \\
5205.724 &       & 1.033 &  -0.340 & 42.89 &  -     & -      &      -     & -     & -        \\
5289.815 &       & 1.033 &  -1.850 & 27.11 &  -     & -      &      33.81 & -     & -        \\
5544.611 &       & 1.738 &  -1.090 & -     &  -     & -      &      36.53 & -     & -       \\ 

\hline
\end{tabular}

This table is available in its entirety in online only. A portion is shown here for guidance regarding its form and content.\\ 
\end{table*}
}
{\footnotesize
\begin{table*}
{\bf  Table 9B: Equivalent widths in m{\rm \AA} of lines used for  abundance determination  of heavy elements for the next 6 objects }\\
\begin{tabular}{lllcllllll}
\hline
Wavelength({\rm \AA})&    Element    &    E$_{low}$(ev) &   log gf&  HD~122202&HD~126681&HD~148897 & HD~164922& HD~167768 &HD~204613\\

\hline
4607.327 & Sr I  & 0.000 &  -0.570 &  -     & -    &  89.9  & 58.5 &  68.6     &   77.31 \\    	  
4854.863 &Y II   & 0.992 &  -0.380 &  -     & -    &  98.6  & 53.8 &  72.2     &  82.63  \\    
4883.685 &       & 1.084 &   0.070 &  129.5 & 28.6 &  114.5 & -    &  80.6     &  101.4  \\                      
5087.416 &       & 1.080 &  -0.170 &  -     & 23.8 &  82.3  & 40.5 &  66.1     &  86.53  \\    
5119.112 &       & 0.992 &  -1.360 &  -     & -    &  41.6  & 19.0 &  16.2     &  47.09  \\    
5205.724 &       & 1.033 &  -0.340 &  -     & -    &  -     & -	   &  -        &  92.26  \\    
5289.815 &       & 1.033 &  -1.850 &  -     & -    &  -     & -	   &  14.2     &  15.22  \\    
5544.611 &       & 1.738 &  -1.090 &  -     & -    &  18.6  & -	   &  11.7     &  29.54  \\    

\hline
\end{tabular}

This table is available in its entirety in online only. A portion is shown here for guidance regarding its form and content.\\ 
\end{table*}
}
\section{Discussion on individual stars}
{\bf HD~55496}: Bond (1974) has classified this high velocity object as a 
sub-giant CH star. MacConnell (1972) included this in the category of weak 
lined metal-deficient Ba II star. Being a high velocity object with lower 
metallicity, ([Fe/H] = $-$1.45)  HD~55496 seems to show the extreme halo 
kinematics. Luck and Bond (1991) has studied this object and reported
 abundances for a few elements  (Table 11). Estimated Ba abundance 
([Ba/Fe] = 0.57)  does not qualify the object to be a typical CH star. Light 
s-process elements Sr, Y and Zr are  more abundant in this star than the 
heavy s-process elements Ba, Ce, and Pr.

{\bf HD~89668}: We  present  first time  detailed abundances for this object.
 This object shows large enhancements in La, Ce, Pr, Nd and Sm with [X/Fe] 
values $\ge$ 1; however,  Ba is slightly underabundant with
 [Ba/Fe] ${\sim}$ $-$0.24.

{\bf HD~92545, HD~107574}: North and Duquennoy (1991) {\bf have} categorized these 
objects as F str Lambda 4077 stars following the classification of Bidelman 
(1981). Allen and Barbuy (2006a, 2006b) have reported  detailed chemical 
abundances 
for these objects  (Table 11).  For HD~92545, our Ce abundance is higher 
than their estimates. Other elements {\bf show} a close similarity  and within the 
error limits. For HD~107574, our results are fairly in good  agreement with 
their estimates.

{\bf  HD~104979, HD~148897}: Luck (1991), identified these objects as cyanogen 
weak giants and reported  elemental abundances for Y, Zr, Ce, Nd and Eu. Our 
results closely match  with their values. In addition to these elements
 we could measure  abundances for Sr, Ba, La, Sm, Pr and Dy. Similar to the 
two cyanogen weak giants HD~188650 and HD~214714 from our paper I, the object 
HD~148897 also does not show large enhancement in heavy elements. These 
three objects are  of the same  spectral type. The object HD~104979 
shows enhancements in Ba  with [Ba/Fe] = 0.94.  
Estimated metallicities of these objects are in the  range  $-$0.2 to $-$1.2. 

{\bf HD~111721}: Gratton and Sneden (1994) have studied this object and 
reported  abundances for heavy elements. From our analysis and also from 
Gratton and Sneden (1994) this object does not show enhancement in 
heavy elements.  The metallicity of this object is = $-$1.11.
This object could be a possible member of the group of CEMP-no stars
of  Beers and Christlieb's (2005) carbon star classification scheme.

{\bf HD~122202, HD~204613}: These two objects are  CH sub-giants. Luck and 
Bond (1991) have studied the object HD~122202 and reported  abundances for 
a few s-process elements. HD~204613 was studied by Smith (1984); these
authors gave  the abundances for Y, Zr, Ba and Nd in this object. In addition 
to these elements we 
estimated the abundances for  Sr, La, Ce, Pr, Sm, Eu and Dy in HD~204613 
and La, Pr and Sm in HD~122202. The object HD~122202  shows a large 
enhancement in Ce, Pr and Nd. However,  Ba is only  mildly enhanced with 
[Ba/Fe] ${\sim}$ 0.33. HD~204613 shows a large enhancement in all the 
elements except Eu. According to Beers and Christlieb (2005) classification,
 this object fall in to the group of CEMP-s stars with [Ba/Fe] ${\sim}$ 1.04
 and [Ba/Eu] ${\sim}$ 0.98.   McClure (1997) have  confirmed these objects 
as binaries. Information on  radial velocity 
variability  and orbital elements for these objects are available in 
McClure (1997). While  HD~122202 shows radial velocity variations in the 
range $-$14.81 to $-$7.64 with an orbital period of 1290 $\pm$ 9 days;
 HD~204613 exhibits radial velocity  variations from -95.07 to -87.85 
with period 878 $\pm$ 4 days.
 
{\bf HD~126681}: We have presented the first time abundance estimates for 
the elements  Ce, Nd and Sm  in this object. Fulbright (2000) has studied 
this object and reported  abundances for Y and Ba.  This object shows 
a large enhancement in Nd and Sm but other heavy elements are only mildly 
enhanced.  

{\bf HD~164922}: The object HD~164922 is listed as a CH star by many authors,
however,  this object 
does not seem to show any characteristics of CH stars. Mishenina et al. (2013) 
have studied this object and reported  abundances for a few heavy elements
that show almost near-solar values for Zr, Ba, Ce,  Nd, Sm and Eu.
Our estimated Ba and Ce abundances give [Ba/Fe] ${\sim}$ 0.28 and [Ce/Fe]
${\sim}$ $-$0.09 for this object.

{\bf HD~167768}: Luck and Heiter (2007) has studied this
object and reported  abundances for Y, Ba, Ce, Pr,
Nd, Eu. Along with these elements we have estimated  abundances 
for Sr, Zr, La and Sm. This object does not show large enhancement
of heavy elements, a  characteristic of CH stars.\\

{\footnotesize
\begin{table*}
{\bf Table 10. Atmospheric parameters from  literature }\\

\begin{tabular}{lccccc}
\hline

Star name &  Vmag  &  T$_{eff}$ (K)  &  log g & [Fe/H]  &  Reference\\
\hline
HD~55496  &8.40  &4850&2.05  &-1.45&1\\
&&4858& 2.05 &-1.48 & 2\\
        & & 4935& 2.33& -1.44&  3\\
        & & 4800 & 2.8 &-1.55 & 4\\
HD~89968&9.41&5400&4.35&-0.13&1\\   
 &  &4811 &4.45 &-0.11&  5\\
HD~92545&8.56&6380&4.65&-0.21&1\\
 &&6240& 4.23& -0.26&  6\\
HD~104979&4.13   &5060&2.67   &-0.26  &1    \\
& &4842 & 2.9& -0.51 & 7\\
        & & 4996& 2.86 &-0.33&  8\\
        & & 4825& 2.34 &-0.33&  9\\
        & & 4870& 3.23 &-0.51& 10\\
        & & 4893&  2.6& -0.29& 11\\
       &  & 4990 &2.65 &-0.11 &12\\
       &  & 5250& 3.25 &-0.29& 13\\
HD~107544&8.55   &6250	& 2.9 & -0.65 &  1  \\
&  &6340& 3.87& -0.36 & 6\\
HD~111721&7.97   &5212&2.6  &-1.11  &1   \\
& &5120& 2.90 &-1.27  &3\\
        & & 4995& 2.52& -1.26& 14\\
        & & 4825  &2.2& -1.54 &15\\
        & & 4800& 3.00 &-1.68& 16\\
       &  & 5164& 3.27& -0.98& 17\\
       &  & 4940& 2.40& -1.34& 18\\
       &  & 5103& 3.06& -1.22& 19\\
        &&  5000&      &-1.34& 20\\
       &  & 5103& 2.87 &-1.25& 21\\
HD~122202&9.37   &6430&4.0&-0.63&1\\
&&6600 & 3.0 &-0.09 & 4\\
HD~126681&9.32   &5760&4.65&-0.90&1\\
&  &5507& 4.45& -1.17& 22\\
       &  & 5561 &4.71 &-1.14 & 5\\
       &  & 5577 &4.25 &-1.12&  2\\
       &&   5475 &4.65 &-1.38& 23\\
       & &  5533 &4.28 &-1.14 &24\\
       & &  5450 & 4.5 &-1.25& 15\\
       &&   5595& 4.43& -1.12 &22\\
        & & 5625& 4.95& -1.09 &17\\
        & & 5500& 4.63& -1.45 &25\\
HD~148897&5.25   &4285&0.6  &-1.02  &1   \\
&&4293 &1.01 &-1.11  &2\\
        & & 4100 &0.09 &-1.16  &4\\
        & & 4345 & 1.5& -0.62 &26\\
HD~167768&6.00   &5070&2.55  &-0.51  &1   \\
&&4953 &2.29 &-0.69 & 2\\
        &  &5102& 2.76 &-0.61 & 8\\
HD~204613&8.21   &5875  &4.2    &-0.24  &1  \\
&& 5718& 3.88 &-0.38&  2\\
        & & 5650& 3.80& -0.35& 27\\
        & & 5650& 3.80 &-0.35& 28\\
        & & 5600&  3.5& -0.70& 29\\
        & & 5600&  3.5& -0.65& 30\\
        & & 5663& 3.75 &-0.54& 31\\

\hline
\end{tabular}

{\bf References.}1. Our work 2. Prugniel et al. 2011, 3. Koleva $\&$ Vazdekis 2012, 4. Luck $\&$ Bond 1991, 5. Sousa et al. 2011,
 6. North et al. 1994 , 7. Massarotti et al. 2008, 8. Luck $\&$ Heiter  2007, 9. Luck 1991, 10. McWilliam 1990, 11. Tomkin $\&$ Lambert 1986, 12. Sneden et al. 1981, 13. Lambert $\&$ Ries 1981, 14. Gratton et al. 2000, 15. Fulbright 2000,
16. Cavallo et al. 1997, 17. Gratton et al. 1996, 18. Ryan $\& $ Lambert 1995, 19. Gratton $\&$ Sneden 1994, 20. Pilachowski et al. 1993, 21.
Gratton $\&$ Sneden 1991, 22. Nissen $\&$ Schuster 2011, 23. Sozzetti et al.2009, 24. Nissen et al. 2000, 25. Tomkin et al. 1992, 26. Kyrolainen
    et al. 1986, 27. Frasca et al. 2009, 28. Smith et al. 1993, 29. Rebolo et al. 1988, 30. Abia et al. 1988, 31. Smith $\&$ Lambert 1986.

\end{table*}
}
{\footnotesize
\begin{table*}\tiny
{\bf  Table 11: Comparison of our results with literature values}\\
\begin{tabular}{lllllllllllll}
\hline
Star Name& [Sr I/Fe]&[Y II/Fe]&[Zr II/Fe]&[Ba II/Fe]&[La II/Fe]&[Ce II/Fe]&[Pr II/Fe]&[Nd II/Fe]&[Sm II/Fe]&[Eu II/Fe]&[Dy II/Fe]&References\\
\hline
Sub giant-CH stars\\
HD~122202 & -  &  1.44& -    & 0.33  &0.9  & 1.62 & 1.26 &-  &  1.77 &-  &  -&1\\
 & -  &  0.74& 0.15   & -  &0.79  & 0.74& - &0.75  &  0.44 &-  &  -&2\\
HD~204613 & 1.71& 0.97& 1.14&  1.04 & 1.21 & 1.24 & 1.52& 1.02& 1.61& 0.06 &1.77&1\\
& -& 1.22& 1.00&  0.71& - & - & -& 0.77& -& -&-&3\\
$\#$CH stars\\
$^∗$HD~ 55496& 0.82& 0.85& 0.52  &0.57 & -  &   0.13 & 0.43& -   & -   & -   & -&1\\
$^∗$& - & -& 0.72  &- & 0.52 &   0.32 & -& 0.52   & 0.33   & -   & -&2\\
HD~89968 &  1.06& 0.55& -  &   -0.24 &1.87 & 1.52 & 1.66 &1.44& 1.23& 0.38 &-&1\\
HD~92545&   -    &0.23& -    & 0.91 & 0.95 & 1.6 &  -   & -   & -  &  -  &  -&1\\
&   0.67    &0.64& 0.75    & 1.04 & 0.72 & 0.60 &  0.44   & 0.42   & 0.24  &  0.32  &  0.09 &4\\
HD~104979 & 0.99& 0.71& 0.85 & 0.94  &1.11 & 1.06  &1.04 &1.13 &1.17 &0.40 &-&1\\
 & -& 0.52& 0.40 & -  &-& 0.48  &- &0.82 &0.51& 0.61&-&5\\
HD~107574 & -   & 1.02& -    & 0.97 & 1.04 & 0.6  & -   & - &   -   & - &   -&1\\
 & 0.67   & 0.64& 0.75    & 1.04 & 0.72 & 0.6  & 0.44   & 0.42 &   0.24   & 0.14 &   0.09&4\\
HD~111721 & -   & 0.05 &-    & -0.09 &0.31&  1.6&   -   & 2.1 & -   & -  &  -&1\\
 & -0.13   & 0.14 &-0.45  & 0.08 &0.04&  0.03&   0.28   & 0.17 & 0.22   & 0.36  &  -& 6\\
&  -   & 0.02 &-    & 0.27 & -    & 0.67 & -   & 1.2 & 1.07 &-   & -&1\\
HD~126681&  -   & 0.23 &-    & 0.14 & -    & - & -   &- & - &-   & -&7\\
HD~148897&  0.31& 0.03 &-0.47& -0.65& 0.29  &-0.16& -   & 0.13 &0.58& -   & 0.02&1\\
&  -& 0.04 &-0.29& -& -  &0.07& -   & 0.01 &-& -0.13   & -& 5\\
HD~164922&0.79 & 0.14 &-&  0.28 &-&  -0.09&- &-&-& -& -&1 \\
         & -& -0.15&0.04 &-0.10 &- & 0.08 &-	&0.03 &	0.01 &	0.10&-\\
HD~167768&  0.77& 0.56& 0.2&   -0.36 &-0.54 &0.06 & -    &0.65 &0.9 & 0.26 &1.04&1\\
&  -& 0.03& -&   -0.03 &- &-0.04 & -0.10    &-0.14&- & 0.45 &-&8\\
\hline
\end{tabular}
$^\#$ Objects from the CH star catalogue of Bartkevicius (1996)\\
$^*$ Objects are also included in Ba star catalogue of L\"u (1991)\\
1. Our work  2. Luck \& Bond (1991) 3. Smith et al. (1993) 4. Allen and Barbuy (2006a) 5. Luck 1991
6. Gratton \& Sneden (1994) 7. Fulbright (2000) 8. Luck \& Heiter (2007)
\end{table*}
}

\section{Stellar masses}
We could estimate the stellar  masses  for eight objects in our sample  
from their locations in
the Hertzsprung-Russel diagram  (Figures 8 and 9), using  
the  evolutionary tracks (Girardi et al. 2000) in the mass range of 
0.15 M$_{\odot}$ to 7.0 M$_{\odot}$ and the Z values from 0.0004 to 0.03 
These evolutionary tracks are available at http://pleiadi.pd.astro.it/. 
 For the objects with near-solar metallicity 
 we have selected an initial composition of Z=0.0198, Y=0.273. 
 The masses derived using spectroscopic temperature estimates are presented  
in Table 12.  
For six stars in our sample that have metallicities $<$ $-$0.5 we 
 used the evolutionary tracks corresponding to 
 Z = 0.008. It is to be noted that   the values of the  masses obtained for 
these objects  with Z = 0.008  are found to be  similar to those 
obtained using evolutionary tracks corresponding to  Z = 0.019. 
Derived stellar masses  are in the range  0.6 M$_{\odot}$ to 1.6 M$_{\odot}$ 
with HD~55496 having a mass of 1.6 M$_{\odot}$ and HD~148897 ${\sim}$
 0.6 M$_{\odot}$. Stellar masses could not be estimated for the  rest of 
the objects as the parallax  estimates are not available in the literature.
\begin{figure}
\centering
\includegraphics[height=8cm,width=8cm]{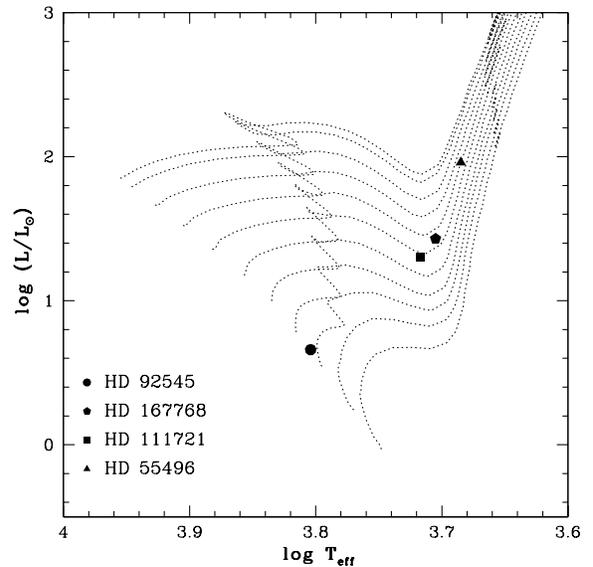}
\caption{The location of HD~92545, HD~167768, HD~111721 and HD~55496 are
indicated  in the H-R diagram. The masses are
derived using the evolutionary tracks of Girardi et al. (2000).
 The evolutionary tracks for masses 1, 1.1, 1.2, 1.3 1.4, 1.5, 
1.6 1.7, 1.8, 1.9 and  1.95 M$_{\odot}$ from bottom to top are shown 
in the  Figure.}
\end{figure}
%label(fig 8)

\begin{figure}
\centering
\includegraphics[height=8cm,width=8cm]{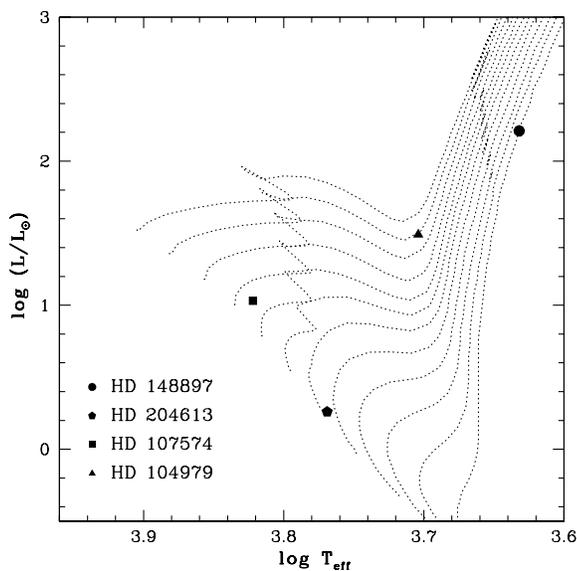}
\caption{
The location of HD~148897, HD~204613, HD~107574 and HD~104979 are 
indicated  in the H-R diagram. The masses are
derived using the evolutionary tracks of Girardi et al. (2000).
The evolutionary tracks are shown for masses 0.6, 0.7, 0.8, 0.9, 1.0,  1.1, 1.2, 1.3 1.4, 1.5, 
1.6 1.7, 1.8, 1.9 and  1.95 M$_{\odot}$ from bottom to top.}
\end{figure}
%\label(fig 9)

\begin{table*}
{\bf  Table 12: Stellar Masses }\\
\begin{tabular}{lccc}
\hline
Star Name.   &$M_v$& $log(L/L_{\odot})$ & $Mass(M_{\odot})$  \\
\hline
HD~55496   &-0.16& 1.96  & 1.6\\
HD~92545  & 3.1&0.66  & 1.2 \\  
HD~104979& 0.63 &1.49 &  1.6 \\
HD~107574 &2.1 &1.03  & 1.45\\
HD~111721 & 1.2&1.3   & 1.5  \\ 
HD~148897 &2.3 &2.21  & 0.60 \\
HD~167768&2.1  &1.43 &  1.55 \\ 
HD~204613 &3.9& 0.26 &  1.1\\

\hline-
\end{tabular}
\end{table*}

\section{Parametric model based study}
Elements heavier than iron are mainly produced by 
two neutron-capture processes, the s-process and the r-process. 
Observed abundances of heavy elements estimated using model atmospheres and
spectral synthesis techniques do not provide direct quantitative estimates 
of the relative contributions from s- and/or r- process nucleosynthesis. 
Identification
of the dominant processes contributing to the heavy element abundances
in the stars is likely to provide clues to their origin.
We have investigated ways to delineate the observed abundances into  
their respective r- and s-process contributions in the framework of a 
parametric model using an appropriate model function.
The origin of the n-capture elements   is explored by comparing the 
observed abundances with predicted s- and r- process contributions 
following Goswami et al. (2010c, and references there in). 
 The ith element abundance can be calculated as 
                        
         $N_i$(Z)  = $A_sN_{is}$  + $ A_r N_{ir}$  10$^{[Fe/H]}$

where  Z is  the metallicity of the star, $N_{is}$  indicates the  abundance  
from s-process in  AGB  star,   $N_{ir}$ indicates the  abundance  
from r-process; $A_s$ indicates  the component coefficient that 
correspond to   contributions   from  the s-process  and
$A_r$ indicates  the component coefficient that correspond to contributions 
  from the r-process.
 
We have utilized the solar system s- and r-process isotopic abundances 
from stellar models of  Arlandini et al. (1999). The observed elemental 
abundances are scaled to the metallicity of the corresponding CH star 
and are normalised to their respective Ba abundances. Elemental abundances 
are then fitted with the parametric model function. The best fit coefficients 
and reduced chi-square values for a set of CH stars are given in Table 13. 
 The best fits obtained with the parametric model 
function $log{\epsilon_i}$  = $A_sN_{is}$  + $ A_r N_{ir}$ for HD~92545, 
HD~104979, HD~107574 and  HD~204613 are shown in Figures 10 -13.
The errors in the derived abundances play an important role in deciding 
the goodness of fit of the  parametric model functions. 
From the parametric model based analysis we find  the objects 
HD~92545, HD~104979, HD~107574 and  HD~204613 to 
 belong to the group of CEMP-s stars.  
                                       
\begin{table*}
{\bf  Table 13: Best fit coefficients and reduced chi-square values}\\
\begin{tabular}{lccc}
\hline
Star Name& $A_s $& $A_r$&$ {\chi}^2$\\
\hline
HD~92545 & 0.560$\pm$ 0.33 &0.503  $\pm$0.33& 2.15\\
HD~104979& 0.514$\pm$ 0.16& 0.493  $\pm$0.15& 0.50\\
HD~107574& 0.823$\pm$0.01 &0.171  $\pm$0.01& 1.22\\
HD~204613& 0.739$\pm$0.08 &  0.291$\pm$0.08 & 1.65\\

\hline
\end{tabular}
\end{table*}

\begin{figure}
\centering
\includegraphics[height=8cm,width=8cm]{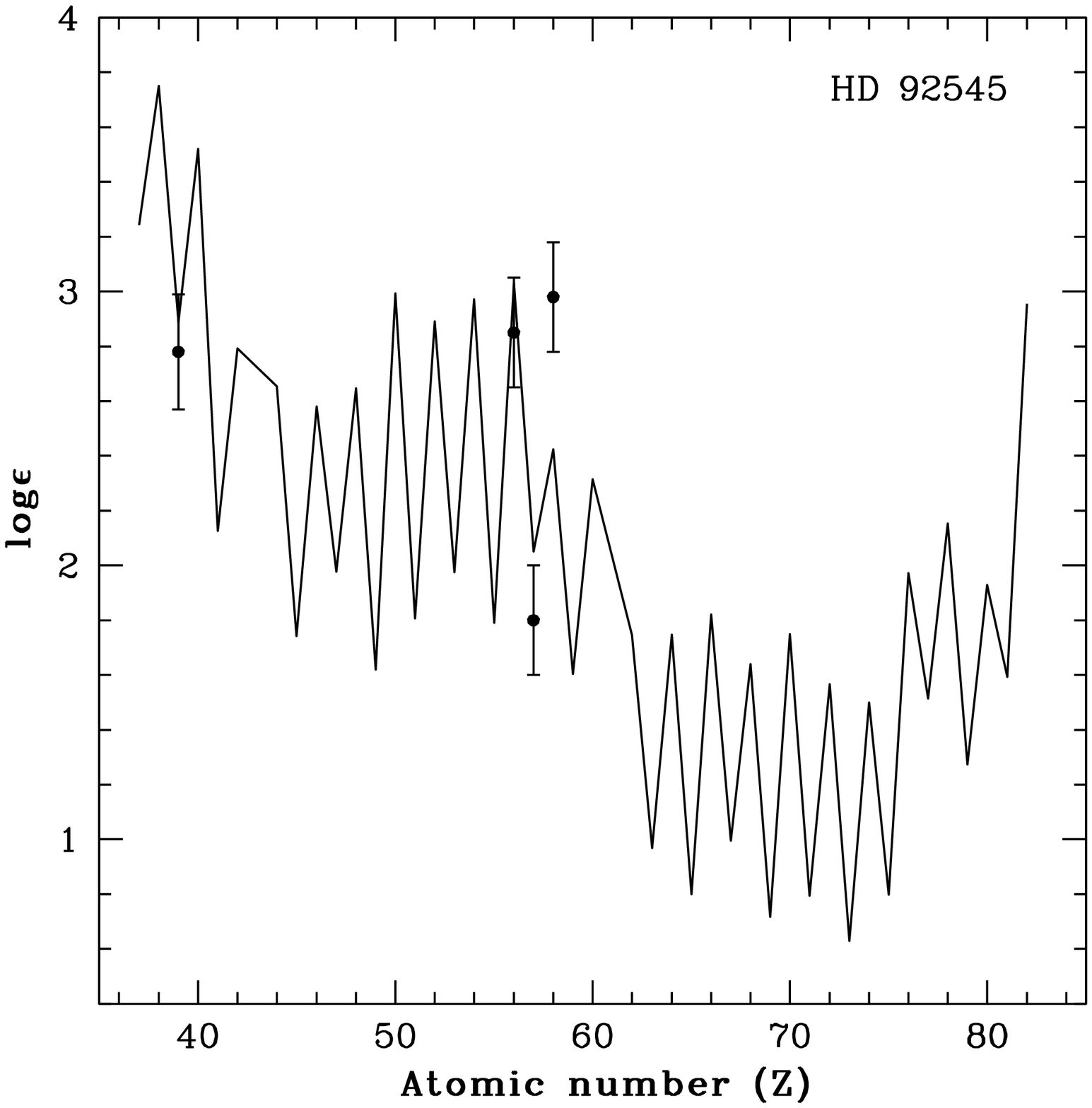}
\caption{  Solid curve represent the best fit for the parametric model
function log${\epsilon}$  = A$_{s}$N$_{si}$ + A$_{r}$ N$_{ri}$, where N$_{si}$
 and N$_{ri}$ represent the abundances
due to s- and r-process respectively (Arlandini et al (1999), Stellar model, scaled
to the metallicity of the star). 
The points with error bars
 indicate the observed abundances in  HD~92545.}
\end{figure}

\begin{figure}
\centering
\includegraphics[height=8cm,width=8cm]{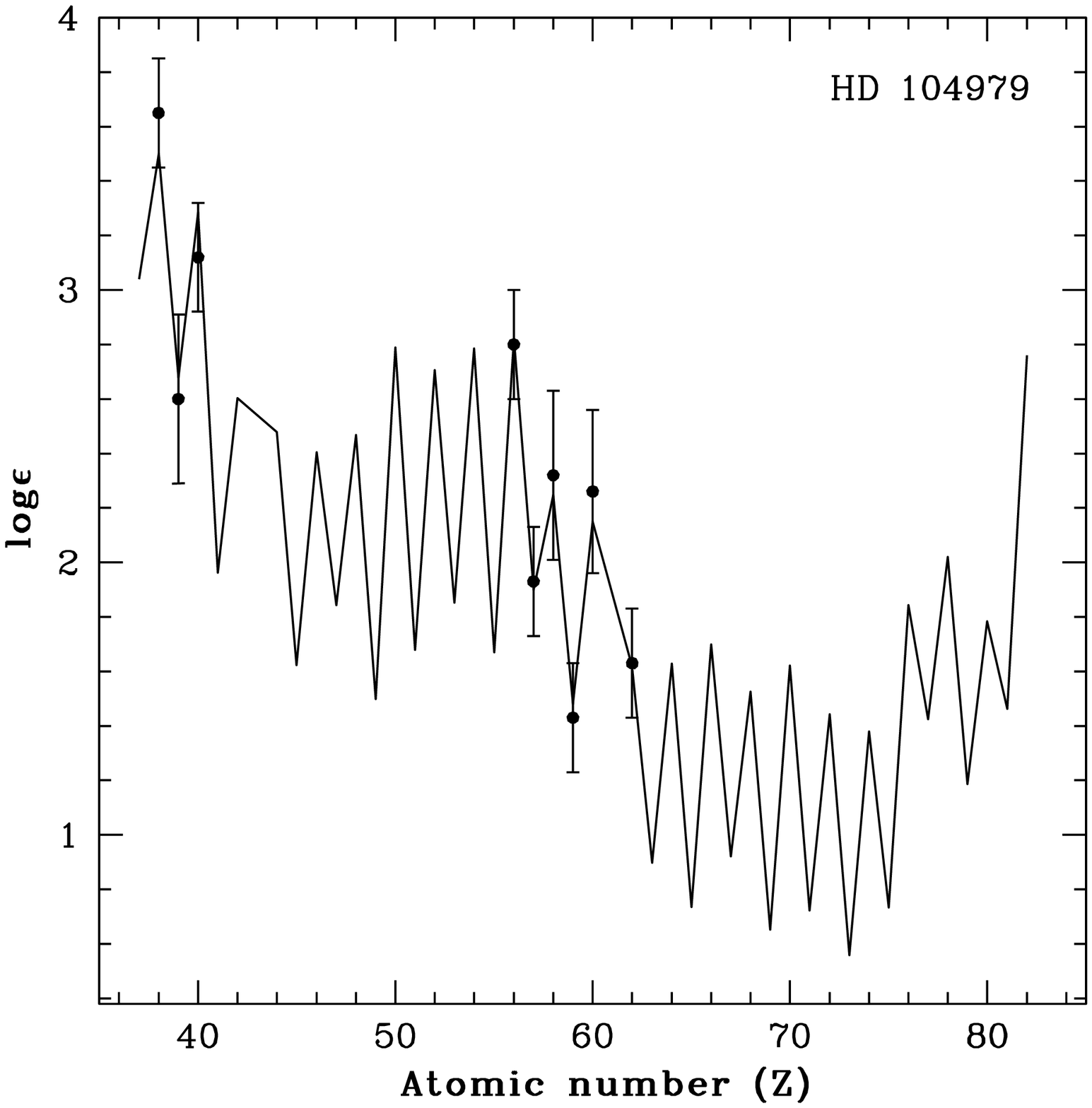}
\caption{  Solid curve represent the best fit for the parametric model
function log${\epsilon}$  = A$_{s}$N$_{si}$ + A$_{r}$ N$_{ri}$, where N$_{si}$
 and N$_{ri}$ represent the abundances
due to s- and r-process respectively (Arlandini et al (1999), Stellar model, scaled
to the metallicity of the star). 
The points with error bars
 indicate the observed abundances in  HD~104979.}
\end{figure}

\begin{figure}
\centering
\includegraphics[height=8cm,width=8cm]{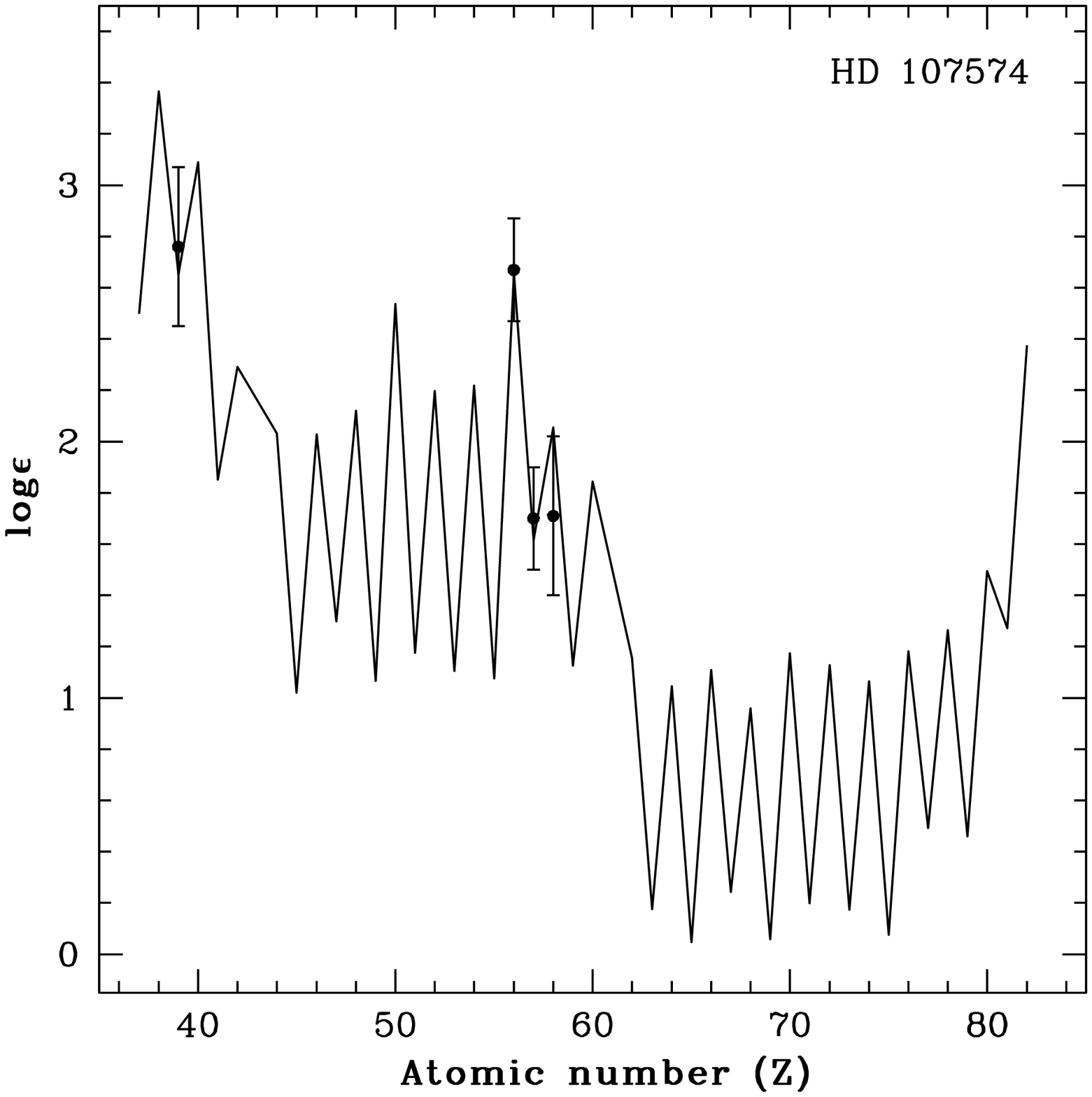}
\caption{  Solid curve represent the best fit for the parametric model
function log${\epsilon}$  = A$_{s}$N$_{si}$ + A$_{r}$ N$_{ri}$, where N$_{si}$
 and N$_{ri}$ represent the abundances
due to s- and r-process respectively (Arlandini et al (1999), Stellar model, scaled
to the metallicity of the star). 
The points with error bars indicate the observed abundances in  HD~107574.}
\end{figure}

\begin{figure}
\centering
\includegraphics[height=8cm,width=8cm]{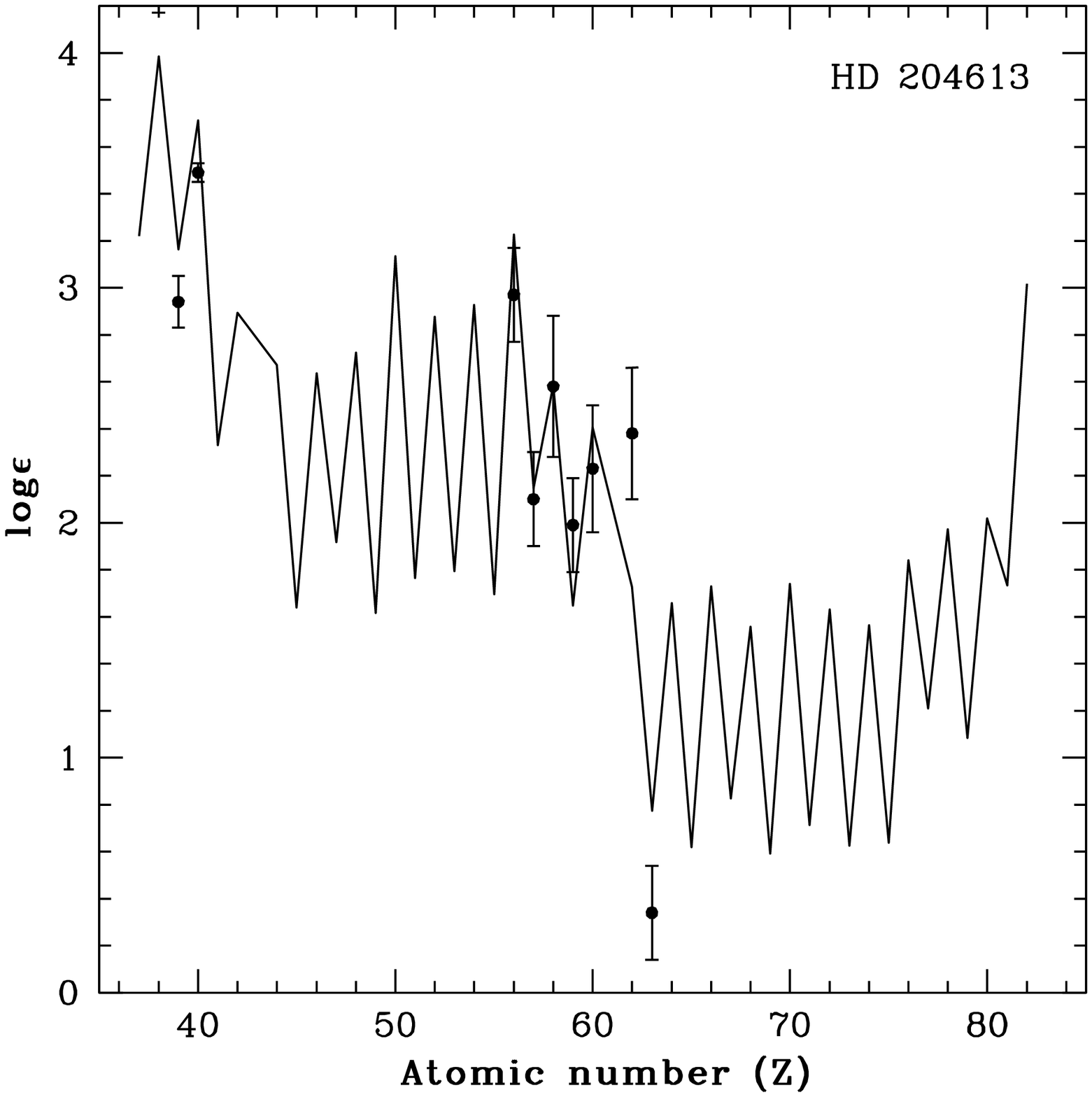}
\caption{  Solid curve represent the best fit for the parametric model
function log${\epsilon}$  = A$_{s}$N$_{si}$ + A$_{r}$ N$_{ri}$, where N$_{si}$
 and N$_{ri}$ represent the abundances
due to s- and r-process respectively (Arlandini et al (1999), Stellar model, scaled
to the metallicity of the star). 
The points with error bars indicate the observed abundances in  HD~204613.}
\end{figure}

\section{Conclusion}
Results from our analyses of a  group of twelve  stars from the CH star 
catalogue of Bartkevicius (1996)  are presented.  Abundances for 22 elements
are estimated.  Except for HD~55496 with radial velocity 315.2 Kms$^{-1}$,
the rest  are low velocity objects. 
HD~55496 is also listed in the Ba star catalogue of  L\"u (1991). 
This object  with a metallicity of $-$1.49 and [C/Fe] ratio of 1.01   shows  a mild enhancement in 
neutron-capture elements. Estimated  [Ba/Fe]  for this object is ${\sim}$ 0.57.

In the sample we have two confirmed binaries; 
HD~122202 and HD~204613 with periods 1290 $\pm$ 9 days and  878 $\pm$ 4 days 
respectively (McClure 1997).
The estimated [C/Fe] is $<$ 1  for all objects  except for HD~55496.
Thus if we follow the CEMP stars classification of Beers and Christlieb (2005)
only HD~55496 falls into  the CEMP star group with [Fe/H] $\le$ -1.0 and
[C/Fe] $\ge$ 1.0. Several authors  have adopted
[C/Fe] ${\ge}$ 0.5  to define CEMP stars (Ryan et al. (2005), Carollo et al. (2012)). In  our sample
four objects have  [C/Fe] ${\ge}$ 0.5.  The  Objects  HD~89668, HD~111721,
HD~148897, HD~164922 and HD~167768 give near solar or mildly under solar
value for  [C/Fe].  These objects also show near solar or underabundant [Ba/Fe]
 value. Although other heavy elements are mildly enhanced in these objects,
these objects are unlikely to belong to the group of  CEMP or classical CH
stars.

We have estimated  the Ba abundance for all the objects in our sample, however
abundance of Eu could be  measured  only for four objects. Following the
abundance criteria of  Beers 
and Christlieb (2005)  based on Ba and Eu abundances two objects 
HD~104979 and HD~204613 with [Ba/Eu] ${\ge}$ 0.5, fall into the group of 
CEMP-s stars. Both  the objects show enhancement in heavy 
elements. In HD~104979 the heavy s-process elements are more enhanced 
than the light s-process elements with [hs/ls] ${\sim}$ 0.18. In HD~204613 
the light s-process elements are more enhanced with [hs/ls] = $-$0.1. 
The parametric model based analysis  indicates  higher contribution 
from the s-process than that of   r-process to the abundances of heavy 
elements observed  in these objects. 

CH stars are  low-mass objects.  Eight objects  in our sample 
for which we could estimate stellar masses 
are found to be low-mass objects with masses in the range 
 0.6 M$_{\odot}$ to 1.6 M$_{\odot}$.
 Stellar masses could not be estimated for the  rest four
 objects as the parallax  estimates are not available in the literature.
These  four objects HD~122202, HD~89968, HD~126681 and HD~164922 have [Ba/Fe]
${\textless}$0.33 with HD~89968 giving a [Ba/Fe] estimate  of ${\sim}$ $-$0.24.
These objects do not  qualify as CH stars.
Abundance ratios of the sample stars show a large scatter with respect to [Fe/H]
(Figure 14). [X/Fe] ratios of the heavy elements for most of the objects
belonging to group 3 are distinctly lower than their counterparts observed in the 
stars of group 1 and 2. Abundance ratios of Eu with respect to Fe observed in
three stars of group 3 show similar values as those seen in two objects of group 2.

Population I Ba stars are believed to be  metal-rich counter parts of CH stars.
 Both CH stars and Ba stars are known to show enhancement in heavy elements.
A comparison of the abundance ratios of  heavy elements with those observed in 
barium stars (solid squares) and CEMP stars from  Masseron et al. (2010) (solid
pentagons) within  the metallicity range 0.2 to  $-$2.2   show that 
the group 3 objects distinctly return lower [X/Fe] ( [Zr/Fe], [Ba/Fe], [La/Fe]
and [Ce/Fe]; Figure 15). These objects do not seem to belong to the group of 
CH stars as far as the chemical composition of heavy elements are concerned.

\begin{figure}
\centering
\includegraphics[height=8cm,width=8cm]{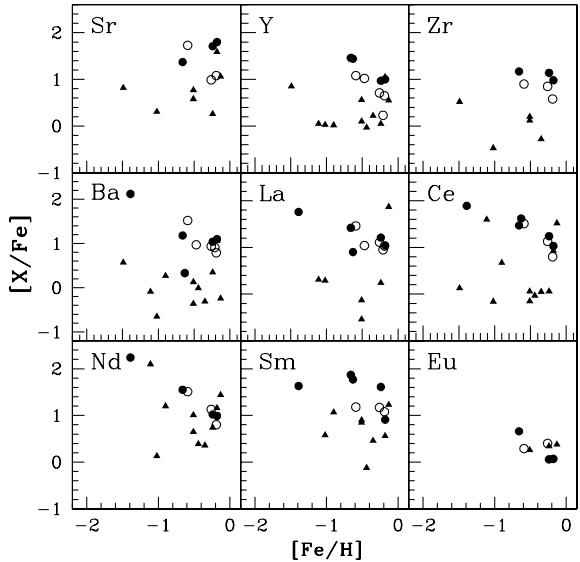}
\caption{  Abundance ratios of heavy elements observed in the program
stars with respect to [Fe/H]. The confirmed binaries are shown with solid circles,
the objects with limited radial velocity information are shown with open circles,
and the rest of the objects are indicated with solid triangles.
  The abundance ratios show a large scatter 
with respect to metallicity.}
\end{figure}
\begin{figure}
\centering
\includegraphics[height=8cm,width=8cm]{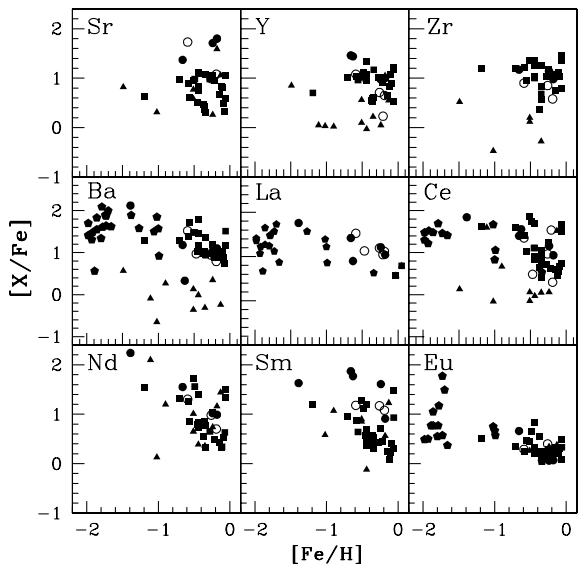}
\caption{ Estimated abundance ratios of Ba, La, Ce and Eu with respect to Fe
  are plotted in this figure where  solid circles indicates the confirmed binaries, 
open circles indicate the objects with limited radial velocity information 
and the solid triangles  indicate the rest of the objects in our sample. The 
abundance  ratios are compared with 
the abundance ratios observed in CEMP stars 
(solid pentagons) from Masseron et al. (2010) and  Ba stars (solid squares) 
from Allen and Barbuy (2006a).}
\end{figure}

{\it Acknowledgement}\\
We thank the referee, for valuable comments which have improved our paper considerably.
This work made use of the SIMBAD astronomical database, operated at CDS, 
Strasbourg, France, and the NASA ADS, USA.  Fundings from   CSIR and DST 
project No. SB/S2/HEP-010/2013 are  gratefully  acknowledged.\\

\end{document}